\documentclass[aps,prc,groupedaddress,twocolumn,showpacs,amsmath,amssymb,floatfix]{revtex4}

\usepackage{graphicx}%
\usepackage{dcolumn}
\usepackage{longtable}
\begin{document}

\title
{Modeling of deuteron-induced reactions on molybdenum at low energies}


\author{M.~Avrigeanu$^{1}$\footnote{marilena.avrigeanu@nipne.ro},
E.~\v Sime\v ckov\'a$^{2}$,
J.~Mr\'azek$^2$,
C.~Costache$^1$, 
V.~Avrigeanu$^1$
}

\affiliation{$^1$Horia Hulubei National Institute for R and D in Physics and Nuclear Engineering, P.O. Box MG-6, 077125 Magurele, Romania}
\affiliation{$^2$Nuclear Physics Institute of the Czech Academy of Sciences (NPI CAS), 25068 \v Re\v z, Czech Republic}


\begin{abstract}
The activities of the EUROfusion consortium on the development of high quality nuclear data for fusion applications include evaluations of deuteron induced reactions and related data libraries for needs of the DEMO fusion power plant and IFMIF-DONES neutron--source nuclear analyses. 
Molybdenum is one of the major constituents of the reference stainless steels used in critical components of these projects.
While the TENDL deuteron data library was the current reference used by EUROfusion, need of its further improvement has already been pointed out. 
The weak binding energy of the deuteron is responsible for the high complexity of its interaction with nuclei, involving also a variety of reactions initiated by the nucleons following the deuteron breakup. 
Their analysis completed that of the deuteron interactions with Mo and its stable isotopes, from elastic scattering to pre-equilibrium and compound--nucleus reactions, up to 50 MeV. 
A particular attention has been paid to the breakup, stripping, and pick-up direct interactions which amount to around half of the deuteron total--reaction cross section.
The due account of most experimental data has validated the present approach, highlighted some prevalent features, and emphasized weak points and consequently the need for modeling/evaluation upgrade. 
\end{abstract}




\maketitle

\section{Introduction}\label{Sec1}

Accurate cross--section data of deuteron induced reactions are a pre--requisite also for  reliable design of the accelerator--based neutron source facility IFMIF-DONES (International Fusion Materials Irradiation Facility – DEMO Oriented NEutron Source), e.g. \cite{db22,wk21} and Refs. therein, and material qualification of the European DEMO fusion reactor \cite{gf17} beyond the ITER fusion--device materials \cite{lwp18,lwp21}. 
Actually, the suitable account of deuteron-nucleus interactions is an important test for reaction mechanism models and nuclear--data evaluation within on-going strategic research programmes at large-scale facilities using deuteron beams \cite{uf20,uf17,nfs,saraf,nfs2021}. 
The present work concerns in this respect the deuteron--induced reactions on Mo, which 
is used in stainless steels to provide a greater corrosion resistance \cite{ap08,xw22}. 
SS-316L steel, being a reference material for Li loop piping of DONES contains 2-3\% of Mo \cite{dones}. 
Since the natural Mo consists of seven stable isotopes, with abundances between 9-24\%, the activation of number of enriched foils in separate experiments would be preferred but for obvious reasons less feasible. 
Thus, a new experiment performed with $^{nat}$Mo targets in Center of Accelerators and Nuclear Analytical Methods (CANAM) infrastructure \cite{CANAM} of the Nuclear Physics Institute of the Czech Academy of Sciences (NPI CAS) and Neutrons for Science Facility at SPIRAL-2 \cite{nfs} was less resource-demanding while the analysis (both experimental \cite{nfs2021,Mod} and theoretical) becomes a quite complex task. 
Apart from the applications in fusion technology, knowledge of activation cross-sections of deuteron-induced reactions on molybdenum is very important also for production of the medically relevant radionuclides (e.g. \cite{tak12,elbinawi,tak11}). 

The deuteron sub-library of the TALYS--based evaluated nuclear data library (TENDL) \cite{T21}, related to the output of the TALYS nuclear model code system \cite{TALYS}, is the current reference within the EUROfusion Consortium \cite{eurofusion} project of the  Early Neutron Source (ENS) \cite{ai19}. 
However, the need of its further improvement has been pointed out \cite{uf20}. 
The latest version of the Fusion Evaluated Nuclear Data Library (FENDL) \cite{FENDL0,FENDL} also includes for structural materials \cite{dl19} the data  of the first version of FENDL, which are from TENDL-2011. 
Meanwhile, recent advancements in deuteron reaction modeling in the TALYS code \cite{uf18,BF_talys} are taken into account to provide more reliable data for TENDL evaluated files, to be reviewed and integrated step by step in the future \cite{FENDL2023}. 

The recent studies of deuteron activation of elements within the priority list of the structural materials \cite{uf09,ra22} known to be problematic (e.g. \cite{ps17}), namely Al, V, Cr, Mn, Fe, Co, Ni, Cu, Zr, and Nb \cite{Ald,VCod,Crd,Mnd,Fed,Nid,Cud,Zrd,Nbd}, are continued by the present work on Mo. 
These nuclear--model analyses of deuteron--induced reactions have been completed with reference to the short reaction--times direct interactions (DI), i.e. breakup (BU) and direct reactions (DR), in addition to the statistical processes of pre-equilibrium emission (PE) and evaporation from compound nucleus (CN) at continuously increasing reaction times.  
A careful account of BU and DR contributions seems essential in this respect due to the still existing disagreement between recent measurements and calculated data \cite{tak12,elbinawi}, while due consideration has also been given to the weak points of an eventual evaluation. 
Moreover, the consistent analysis of all available data for competitive reaction channels as well as full stable isotopic chains and even neighboring elements, beyond a particular reaction analysis for one or only a couple of isotopes (e.g. \cite{ak17}), strengthens the assessment of the model approach.


A consistent energy--dependent optical model potential (OMP) for deuterons on Mo isotopes is concerned in Sec.~\ref{omp} using the computer code SCAT2 \cite{SCAT2}. 
Then, theoretical framework of the deuteron BU mechanism (Sec.~\ref{breakup}) and the DR analysis (Sec.~\ref{DR}) involving the computer codes TALYS-1.96 \cite{TALYS} and FRESCO \cite{FRESCO}, respectively, is briefly mentioned. 
The PE and CN mechanism contributions to the population of various residual nuclei (Sec.~\ref{PE+CN}), on the grounds of also  TALYS-1.96, are completing the nuclear model discussion of the present work. 
The measured and calculated deuteron activation cross sections of natural Mo and its stable isotopes, as well as the corresponding evaluated data within the TENDL-2021 library \cite{T21}, are then compared in Sec.~\ref{Activation}. 
Conclusions of this work are given in Sec.~\ref{Sum}. 

\section{Nuclear model framework} \label{models}

The weak binding energy of the deuteron is responsible for the high complexity of its interaction with nuclei, involving also a variety of reactions initiated by the nucleons following the deuteron breakup (breakup nucleons) \cite{BF_talys,breakup0,breakup}. 
Its importance increases with the target-nucleus mass and charge, becoming dominant for heavy target nuclei at incident energies particularly around the Coulomb barrier \cite{Pad,surr}. 
Overall, the accompanying breakup--nucleons induced reactions make substantially different the deuteron projectile among the other incident particles, with proved significant impact even for applications in nuclear transmutation of radioactive waste \cite{hw19,tc23}. 
Additional comments on systematic uncertainties in DR theories used to analyze particularly $(d,p)$ reactions are referred to in Sec.~\ref{DR}.

\begin{figure*} 
\includegraphics[width=0.65\textwidth]{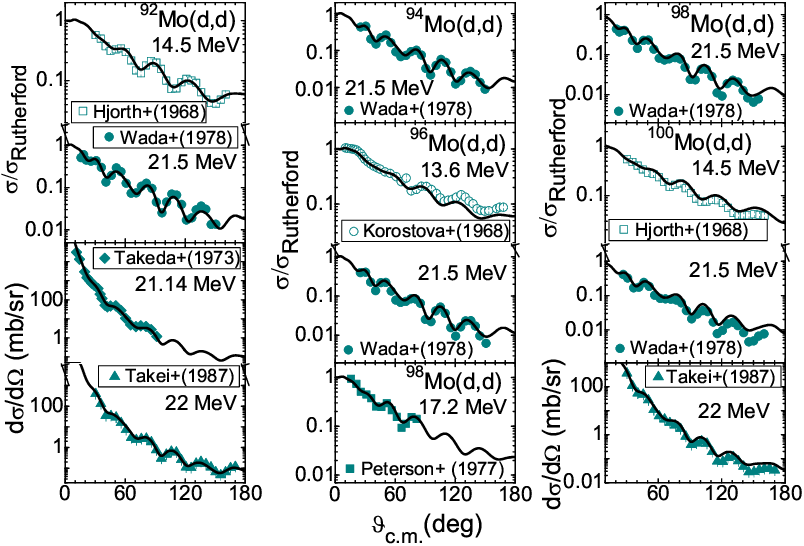}
\caption{(Color online) Comparison of measured \cite{EXFOR,hjorth,korostova,takeda,peterson,wada,takei}
and calculated elastic--scattering angular distributions of deuterons on $^{92,94,96,98,100}$Mo isotopes at 14.5, 17.2, 21.14, 21.5, and 22 MeV, using the global OMP of Daehnick {\it et al.} \cite{dah}}
\label{Mod_ad_el}
\end{figure*}

\subsection{Deuteron optical potential assessment} \label{omp}

A simultaneous analysis of the deuteron elastic scattering as well as induced activation cross sections is essential as the deuteron OMP parameters are obtained by the former data fit, and then used within the analysis of the latter ones. 
The demand of a consistent input of nuclear model calculations of deuteron activation of Al, V, Cr, Fe,  Co, Ni, Zr, Nb, Pa, U \cite{Ald,VCod,Crd,Mnd,Fed,Nid,Cud,Zrd,Nbd,Pad,surr} was satisfied by the deuteron OMP of Daehnick {\it et al.} \cite{dah}. 
It has been a first option of the present work, being obtained by use of a large experimental basis including the angular distributions of elastic scattered deuteron on the above--mentioned elements as well as on $^{100}$Mo isotope. 

\begin{figure}
\centering
\includegraphics[width=0.5\textwidth]{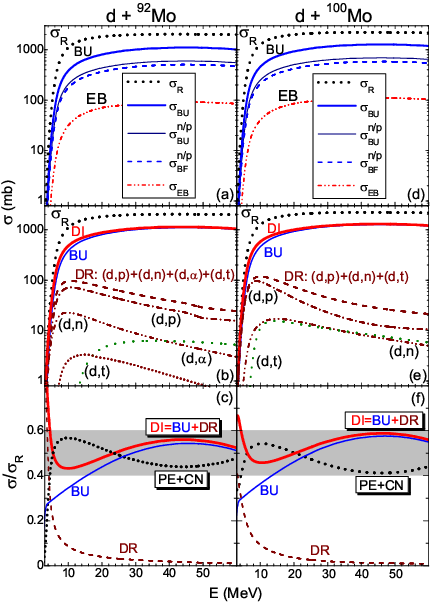}
\caption{(a,d) Excitation functions of deuteron $\sigma_R$ (dotted curves), total breakup $\sigma_{BU}$=2$\sigma^{n/p}_{BF}$+$\sigma_{EB}$ (thick solid curves), $nucleon$-$emission$ total breakup $\sigma^{n/p}_{BU}$=$\sigma^{n/p}_{BF}$+$\sigma_{EB}$ (thin solid curves) and inelastic-breakup $\sigma^{n/p}_{BF}$ (dashed curves), and elastic breakup $\sigma_{EB}$ (dash-dot-dotted curves) cross sections \cite{breakup}; 
(b,e) DI excitation functions (thick solid curves) and its components: total BU (thin curves), DR (dashed curves), stripping $(d,p)$ (dash-dotted curves) and $(d,n)$ (dash-dot-dotted curves), and pick-up $(d,\alpha)$ (short-dotted curve) and $(d,t)$ (dotted curves); 
(c,f) $\sigma_R$ fractions of BU (thin solid curves), DR (dashed curves), DI (solid thick curves), and PE+CN (thick dotted curves) cross sections of deuteron interactions with (a,b,c) $^{92}$Mo and (d,e,f) $^{100}$Mo (see text)}
\label{FR_92Mo-100Mo}
\end{figure}

Unfortunately, the systematics of elastic--scattering angular distributions of deuterons on the stable isotopes of Mo is scarce, despite the number of these isotopes. 
There is also no related measurement of the deuteron total--reaction cross section $\sigma_R$. 
The comparison of the measured angular distributions of elastic--scattered deuterons on $^{92,94,96,98,100}$Mo \cite{EXFOR,hjorth,korostova,takeda,peterson,wada,takei} at the incident energies of 14.5, 17.2, 21.14, 21.5, and 22 MeV, and the calculated values using the computer code SCAT2 \cite{SCAT2} and Daehnick {\it et al.} \cite{dah} OMP is shown in Fig.~\ref{Mod_ad_el}. 
The good description of the measured data provided thus confidence in the further use of this potential within a consistent analysis of the deuteron activation cross sections of  Mo isotopes through the BU, DR, PE, and CN reaction mechanisms.

\subsection{Deuteron breakup}
\label{breakup}

A detailed overview of the specific deuteron breakup in the Coulomb and nuclear fields of the target nucleus, including its complexity given by the addition to the primary deuteron--nucleus interaction of various nuclear reactions initiated by the breakup nucleons, was given recently \cite{BF_talys,Zrd}. 
Therefore, only particular points are mentioned hereafter concerning the two distinct BU processes, i.e. the elastic breakup (EB) in which the target nucleus remains in its ground state and none of the deuteron constituents interacts with it, and the inelastic breakup or breakup fusion (BF), where one of these  constituents interacts nonelastically with the target nucleus. 

Empirical parametrization \cite{BF_talys,breakup00,breakup0,breakup} has concerned the fractions $f_{BU}^{n/p}$ and $f_{EB}$ of cross sections for the total breakup (EB+BF) $nucleon$-$emission$ $\sigma^{n/p}_{BU}$ and elastic breakup $\sigma_{EB}$, respectively, to $\sigma_R$.  
The experimental systematics of deuteron-induced reactions on target nuclei from $^{27}$Al to $^{232}$Th \cite{pamp78,wu79,klein81,mats82,must87} was involved in this respect. 
Moreover, since equal BF nucleon--emission cross sections $\sigma^n_{BF}$ and $\sigma_{BF}^p$ were assumed \cite{must87}, the total breakup  (BF+EB) cross section becomes 
$\sigma_{BU}$=2$\sigma^{n/p}_{BF}$+$\sigma_{EB}$, 
while the total breakup (BF+EB) $nucleon$-$emission$ cross section is 
$\sigma^{n/p}_{BU}$=$\sigma^{n/p}_{BF}$+$\sigma_{EB}$.  
At the same time, the BF fractions for each of the breakup nucleons are given by the difference $f_{BF}^{n/p}$=$f_{BU}^{n/p}$-$f_{EB}$.

An overall view of the incident--energy dependence of the BU, BF, and EB cross sections for deuterons on $^{92,100}$Mo target nuclei is shown in Fig.~\ref{FR_92Mo-100Mo}(a,d). 
It is thus apparent the change of these excitation functions with the target--nucleus mass number, too. 
There are shown also both the BU and BF nucleon--emission excitation functions.    
These results as well as the earlier ones \cite{Ald,VCod,Crd,Mnd,Fed,Nid,Cud,Zrd,Nbd,Pad,surr} point out the dominant role of the BF component, to be concerned in connection with the two opposite effects of the deuteron breakup on the deuteron activation cross sections \cite{BF_talys,breakup0,breakup}. 
At first, the total--reaction cross section shared among various outgoing channels, is reduced by the value of the total breakup cross section $\sigma_{BU}$. 
Next, this component brings contributions to different reaction channels through breakup--neutrons/protons interactions with target nucleus \cite{BF_talys,Ald,VCod,Crd,Mnd,Fed,Nid,Cud,Zrd,Nbd,Pad,surr} leading to the enhancement of the corresponding $(d,xp)$ or $(d,xn)$ reaction cross sections, respectively. 

It should be noted that the BF cross--section partition among various residual--nuclei population is triggered by the energy spectra of the breakup nucleons as well as the excitation functions of the reactions induced by these nucleons on the target nucleus \cite{breakup0,breakup}, while the atomic mass and maybe also the atomic number of the related compound nuclei differ by one unit from that in deuteron-induced reactions. 
The BF related formalism involved in the present work is described in Sec. 3.4 of Ref. \cite{BF_talys}, the corresponding reaction cross sections being calculated with the code TALYS-1.96 \cite{TALYS} and its 
second option for the breakup model, 
i.e. including the BF enhancement (\cite{TALYS}, p. 40).

The enhancements due to $(p,x)$ and $(n,x)$ reactions induced by the breakup nucleons on $^{92,94,96,98,100,nat}$Mo are discussed in Sec.~\ref{Activation} and distinctly showed in  Figs.~\ref{101Tc_natMod}--\ref{89Zr-88Zr_natMod}. 
It has been thus proved that BF enhancements are particularly important for the suitable description of the maximum as well as the high--energy side of the excitation functions for second and third chance emitted-particle channels \cite{BF_talys}. 

\subsection{Direct reactions} 
\label{DR}

Apart from the breakup contributions to deuteron interactions, an increased attention has to be paid to the direct reactions so far poorly considered within deuteron activation analysis. For low-- and medium--mass target nuclei and deuteron energies below and around the Coulomb barrier, the interaction process proceeds largely through DR mechanism, while PE+CN component become important with the incident--energy increase.
Nevertheless, the assessment of DR cross sections is subject to available information on spectroscopic factors for populated states, outgoing particle angular distributions, or at least differential cross--section maximum values. 

The appropriate assessment of stripping -- $(d,p)$, $(d,n)$, and $(^3He,d)$ -- and pick-up -- $(d,t)$, $(d,\alpha)$, and $(d,^3He)$ -- DR mechanism contributions has been performed through the distorted-wave Born approximation (DWBA) method, with the code FRESCO \cite{FRESCO}. 
The $post/prior$ form distorted--wave transition amplitudes for stripping and pick-up reactions, respectively, and the finite--range interaction have been considered. 
A note may concern the finite range (FR) effects in deuteron stripping process which were formerly shown to be very small around the incident energy of 10 MeV, so that a zero--range (ZR) approximation was expected to be reasonable \cite{pjab64} and eventually completed (e.g. \cite{sn21}) by an usual FR correction \cite{xdl04,mbt05}. 
More recently, ZR calculation has led to results very close to the full FR-DWBA calculation for deuteron reactions while FR effects become important for incident $^6$Li ions \cite{jl15}. 
Nevertheless, it was also found these effects are important in $(d,p)$ reactions at intermediate energies (20 MeV/u) \cite{nbn10} while they are not however among the actual major sources of uncertainties in a variety of reaction theories used to analyze $(d,p)$ nuclear reactions in the incident energy range 10-20 MeV \cite{ael15}.
 
At the same time, a simple Gaussian form was considered for the $n$-$p$ interaction in the deuteron \cite{kamimura86}, which reproduces well the binding energy of deuteron and the low--energy $n$-$p$ scattering phase shifts.  
It is yet widely used (e.g. \cite{jl15,jl18}), DWBA as well as BU cross sections being modified by a few percent when including a realistic NN interaction for the reaction at low energy, while at higher energy the cross section is not affect by details of the NN interaction \cite{amm09,nju12,ad16,mgr17,ysn20}. 
Following the original data analysis, similar form has been used for the $d$-$n$ interaction in the triton \cite{triton}, at the same time with a Woods-Saxon shape of the $d$-$p$ interaction \cite{kamimura86} in $^3$He, as well as the $d$-$d$ 
interaction \cite{alpha-d} in the $\alpha$ particle. The transferred--nucleon and --deuteron bound states were generated in a Woods-Saxon real potential \cite{Ald,Fed}. 

The number of $N$ nodes corresponding to an $L$ transferred angular momentum in the radial wave function was determined by the harmonic--oscillator energy conservation rule:

\begin{equation}\label{eq:1}
2N+L=2(n_{n/p}-1)+l_{n/p} \: ,
\end{equation}

\noindent
where $n_{n/p}$ and $l_{n/p}$ are the single--particle (neutron/proton) shell--model state quantum numbers. 
Particularly for the FRESCO code, the number of nodes includes the origin, so that $N$$>$0.

\begin{figure} 
\includegraphics[width=0.48\textwidth]{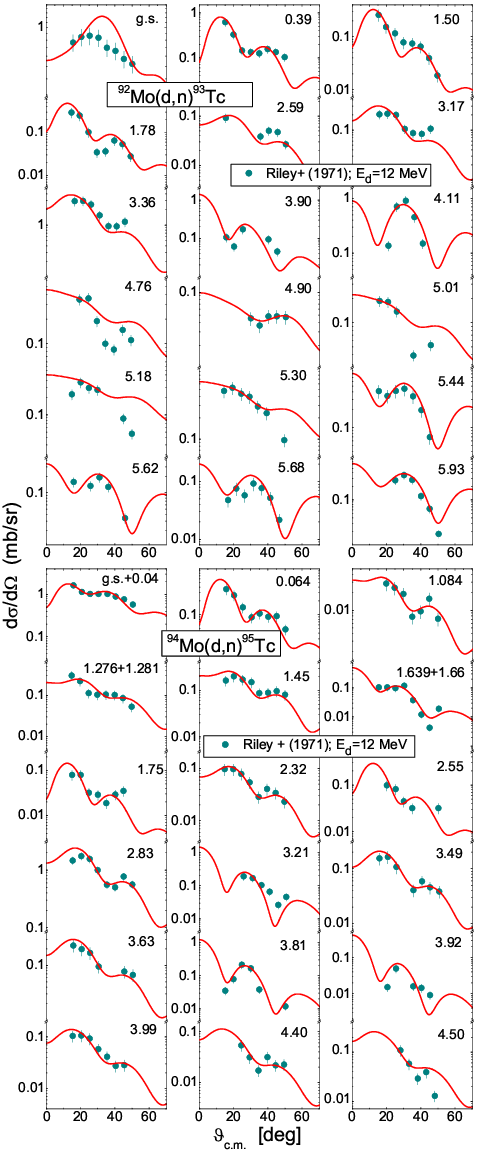}
\caption{(Color online) Comparison of measured \cite{riley} and calculated (solid curves) neutron angular distributions of $^{92}$Mo$(d,n)^{93}$Tc (top) and $^{94}$Mo$(d,n)^{95}$Tc (bottom) stripping transitions to states with excitation energies in MeV, at the incident energy of 12 MeV. 
For the measured angular distributions of unresolved doublets it is shown the sum of the calculated contributions corresponding to the spectroscopic factors for each state of the doublet given in the Appendix~\ref{secA1}.}
\label{92Mo-94Mo_d,n}
\end{figure}

\begin{figure*} 
\includegraphics[width=0.997\textwidth]{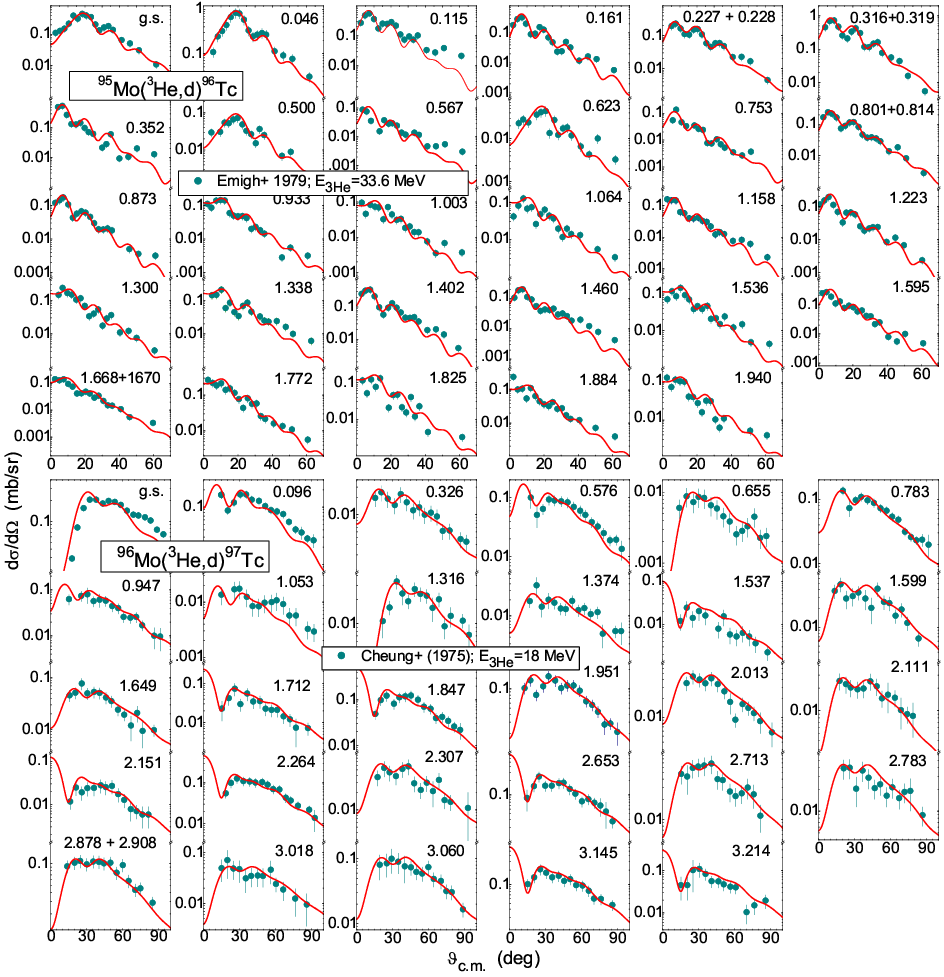}
\caption{(Color online) As Fig.~\ref{92Mo-94Mo_d,n} but for deuteron angular distributions of $^{95}$Mo$(^3He,d)^{96}$Tc \cite{emigh} (top) and $^{96}$Mo$(^3He,d)^{97}$Tc (bottom)  \cite{cheung75} stripping transitions, at 33.6 and 18 MeV, respectively}
\label{95Mo-96Mo_3Hed}
\end{figure*}

\begin{figure} 
\includegraphics[width=0.48\textwidth]{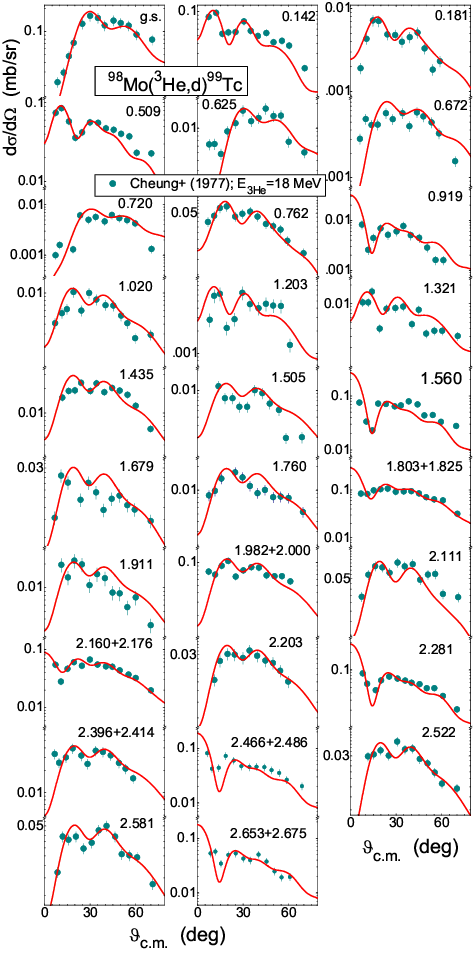}
\caption{(Color online) As Fig.~\ref{95Mo-96Mo_3Hed} but for $^{98}$Mo$(^3He,d)^{99}$Tc \cite{cheung77} at 18 MeV}
\label{98Mo_3Hed}
\end{figure}
 
\begin{figure} 
\includegraphics[width=0.48\textwidth]{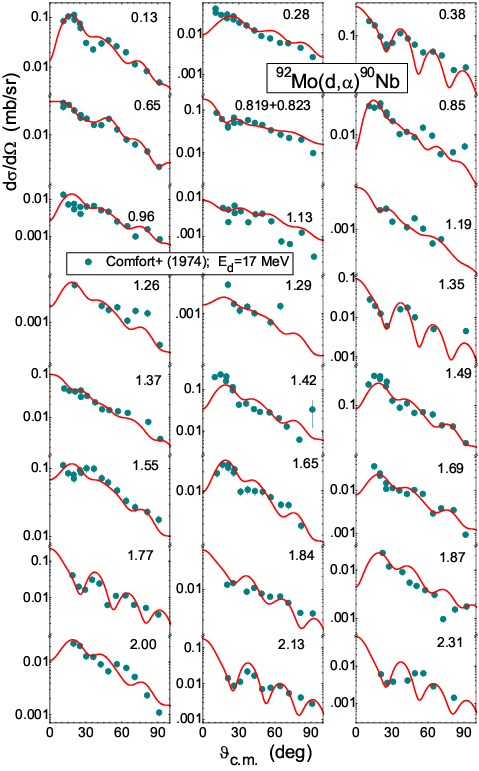}
\caption{(Color online) As Fig.~\ref{92Mo-94Mo_d,n} but for $\alpha$-particle angular distributions of $^{92}$Mo$(d,\alpha)^{90}$Nb pick-up transitions \cite{comfort}, at 17 MeV}
\label{92Mo_d,a}
\end{figure}

\begin{figure*} 
\includegraphics[width=0.997\textwidth]{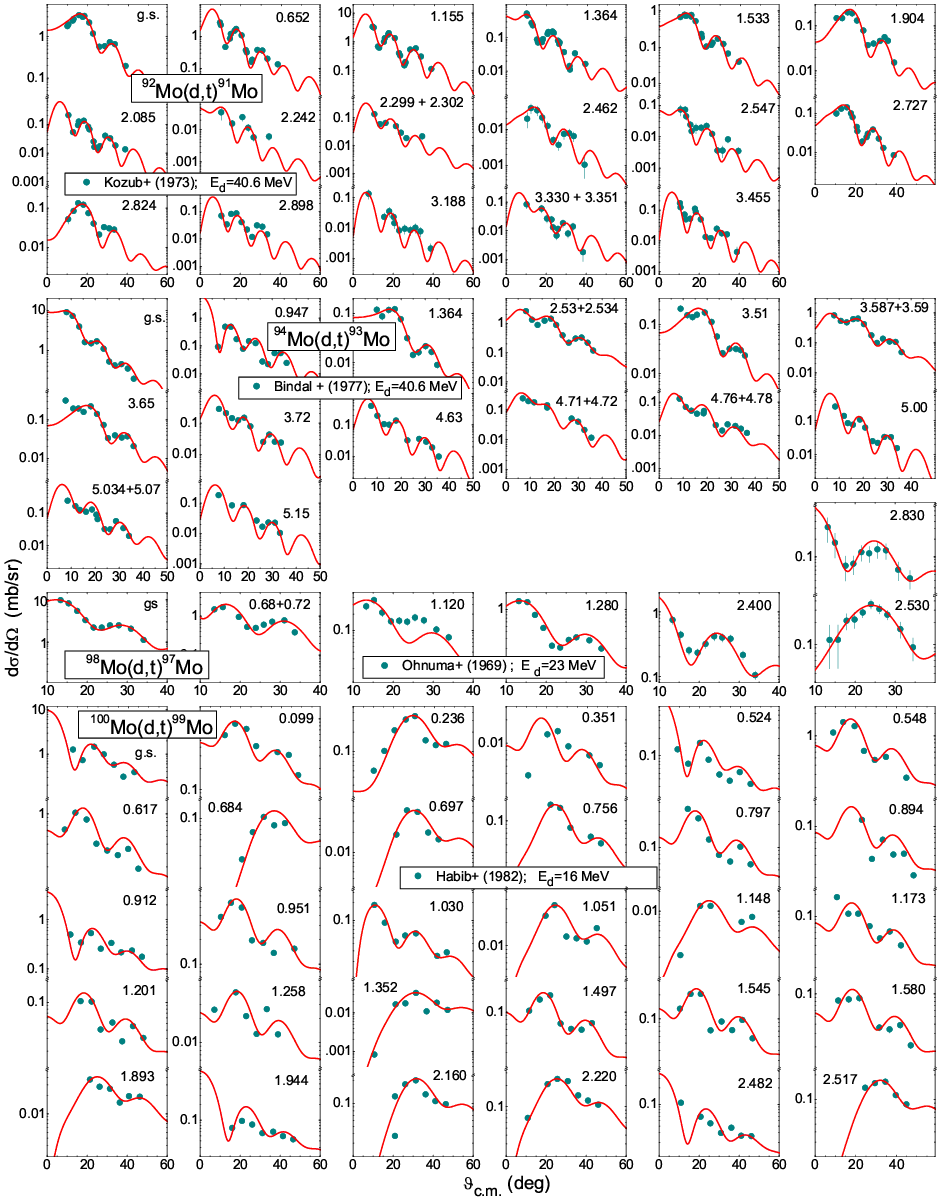}
\caption{(Color online) As Fig.~\ref{92Mo-94Mo_d,n} but for triton angular distributions of $^{92,94,98,100}$Mo$(d,t)^{90,92,94,98,100}$Mo pick-up transitions \cite{kozub,bindalt,ohnuma,habib}, at 40.6 (top), 23 (middle), and 16 (bottom) MeV}
\label{92Mo-94Mo-98Mo-100Mo_dt}
\end{figure*} 

\begin{figure} [!htbp]
\includegraphics[width=0.48\textwidth]{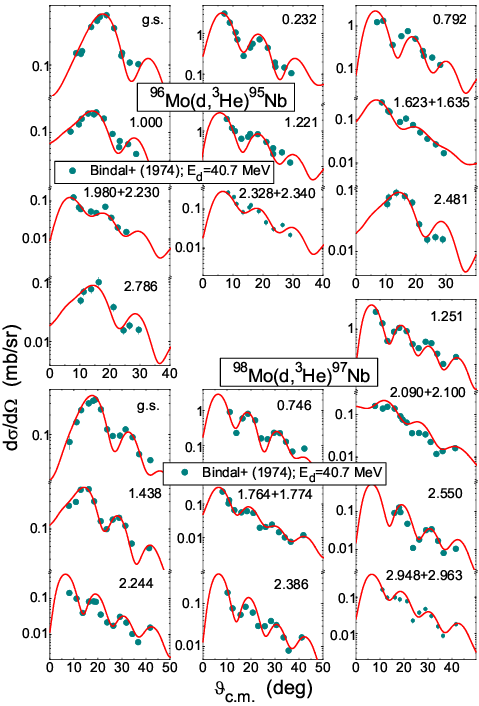}
\caption{(Color online) As Fig.~\ref{92Mo-94Mo_d,n} but for $^3$He angular distributions of $^{96,98}$Mo$(d,^3He)^{95,97}$Nb pick-up transitions \cite{bindal}, at 40.7 MeV}
\label{96Mo-98Mo_d,3He}
\end{figure}

The energy, spin, and parity of the residual--nuclei discrete levels within the ENSDF \cite{BNL} and RIPL \cite{RIPL} libraries were used as starting input of the DWBA calculations for the stripping and pick-up reactions \cite{basu,coral,abriola,nica,browne,mccut,101}.

\subsubsection{Stripping reactions} 

\noindent
{\bf $(d,n)$ reactions.} \\
The analysis of the neutron angular distributions from $^{92,94}$Mo$(d,n)^{93,95}$Tc reactions  \cite{riley} is shown in Fig.~\ref{92Mo-94Mo_d,n} for transitions to discrete levels of residual nuclei. There were thus provided the spectroscopic factors (given in the Appendix~\ref{secA1}) and then the corresponding calculated stripping cross sections, e.g. Fig.~\ref{FR_92Mo-100Mo}(b), Fig.~\ref{95Tc_natMod}(e,j,o), and Fig.~\ref{93Tc_natMod}(d,h,l) in Sec.~\ref{Activation}.  

\bigskip

\noindent
{\bf  $(^3He,d)$ reactions.}\\
There are so scarce or even lacking measured neutron angular distributions for $^{95,96,98}$Mo$(d,n)^{96,97,99}$Tc stripping reactions, in order to obtain the stripped--proton spectroscopic factors for the assessment of the DR contributions
Because of that, these spectroscopic factors have been got through deuteron angular-distribution analysis of $(^3He,d)$ stripping reaction populating the same residual nucleus as the $(d,n)$ process. 
Thus, description of the deuteron angular distributions from $^{95,96,98}$Mo$(^3He,d)^{96,97,99}$Tc stripping reactions \cite{emigh,cheung75,cheung77}  (Figs.~\ref{95Mo-96Mo_3Hed} and ~\ref{98Mo_3Hed}) validates the extracted stripped--proton spectroscopic factors.  Then, they were involved in $^{95,96,98}$Mo$(d,n)^{96,97,99}$Tc stripping cross--sections calculations whose contributions to the total activation cross sections are shown in Figs.~\ref{99mTc_natMod}(b), ~\ref{97mTc_natMod}(b), and ~\ref{96Tc_natMod}(b,g,l)  of Sec.~\ref{Activation}.

Moreover, for calculations of $^{100}$Mo$(d,n)^{101}$Tc stripping cross sections we made use of the spectroscopic factors reported by Freeman {\it et al.} \cite{freeman} (Table XXVII) for deuteron distributions of $^{100}$Mo$(^3He,d)^{101}$Tc stripping reaction. 
The corresponding stripping $^{100}$Mo$(d,n)^{101}$Tc excitation function is shown in Fig.~\ref{FR_92Mo-100Mo}(e) and Fig.~\ref{101Tc_natMod}(b) of Sec.~\ref{Activation}.
The same discrete levels schemes are involved for the residual nuclei within both $(d,n)$ and $(^3He,d)$ reaction calculations, e.g. 33 levels of $^{96}$Tc \cite{BNL,abriola}, 30 levels of $^{96}$Tc \cite{BNL,nica}, 35 levels of $^{99}$Tc \cite{BNL,browne}, and 17 levels of $^{101}$Tc \cite{freeman}.

\bigskip

\noindent
{\bf $(d,p)$ reactions.} \\ The $(d,p)$ stripping reaction cross sections corresponding to the residual nuclei of interest for this work have been calculated using the spectroscopic factors provided by the more recent analyzes of $^{92}$Mo$(d,p)^{93}$Mo \cite{sharp2013} (Table III), and $^{98,100}$Mo$(d,p)^{99,101}$Mo reactions \cite{freeman} (Tables XXIV and XXIII, respectively). 
The related $^{92,100}$Mo$(d,p)^{93,101}$Mo excitation functions are shown in Figs.~\ref{FR_92Mo-100Mo}(b,e) and \ref{101Mo_natMod}(b) of Sec.~\ref{Activation}. 
At the same time, the excitation function of the $^{98}$Mo$(d,p)^{99}$Mo stripping reaction
is shown in Fig.~\ref{99Mo_natMod}(b). 

\subsubsection{Pick-up reactions}

\noindent
{\bf $(d,\alpha)$ reactions.} \\ Even worse than the scarce systematics of neutron angular distributions corresponding to $(d,n)$ stripping processes, there exist measured $\alpha$-particle angular distributions only for the $^{92}Mo(d,\alpha)^{90}$Nb pick-up reaction \cite{comfort}. 
Their analysis shown in Fig.~\ref{92Mo_d,a} has allowed to extract the spectroscopic factors of the picked deuteron. 
The corresponding  $(d,\alpha)$ excitation function is included also in Fig.~\ref{FR_92Mo-100Mo}(b), but particularly in Fig.~\ref{92mNbP-91mNb-90Nb_natMod}(l) of Sec.~\ref{Activation}.  

On the other hand, spectroscopic factors corresponding to  $^{94,97,98}$Mo$(d,\alpha)^{92,95,96}$Nb pick-up reactions have been obtained following due consideration of the measured $\alpha$-particle angular distributions \cite{schoon} maxima. 
The corresponding $(d,\alpha)$ pick-up excitation functions are shown in Figs.~\ref{98mNb-97Nb-96Nb_natMod}(k), ~\ref{95NbP_natMod}(d,i,n), and ~\ref{92mNbP-91mNb-90Nb_natMod}(c) of Sec.~\ref{Activation}.  

\bigskip 

\noindent
{\bf $(d,t)$ reactions.} \\ 
The analysis of measured triton angular distributions \cite{kozub,bindalt,ohnuma,habib} of the $^{92,94,98,100}$Mo$(d,t)^{91,93,97,99}$Mo pick-up reactions has provided the spectroscopic factors for the corresponding transitions to discrete levels of residual nuclei. The suitable account of the available $(d,t)$ angular distributions shown in Fig.~\ref{92Mo-94Mo-98Mo-100Mo_dt} provides the modeling support of the calculated pick-up excitation functions shown in  Figs.~\ref{FR_92Mo-100Mo}(b,e) and \ref{99Mo_natMod}(c). 

\bigskip

\noindent
{\bf $(d$,$^3$$He)$ reactions.}\\ 
The analysis of measured outgoing $^3$He angular distributions of the $^{96,98}$Mo$(d$,$^3$$He)$$^{95,97}$Nb pick-up processes \cite{bindal} made possible further calculations of the corresponding pick-up cross--section contributions to the total activation cross sections of $^{95,97}$Nb residual nuclei. 
Thus, appropriate  description of these angular distributions (Fig.~\ref{96Mo-98Mo_d,3He}) validated the extracted spectroscopic factors which have led to pick-up excitation functions in Figs.~\ref{98mNb-97Nb-96Nb_natMod}(f) and \ref{95NbP_natMod}(c,h,m)  of Sec.~\ref{Activation}. 

\bigskip

The DR contributions, which have proved essential for describing the measured excitation functions corresponding to the first--chance emitted particles \cite{Ald,VCod,Crd,Mnd,Fed,Nid,Cud,Zrd,Nbd}, are furthermore discussed in Sec.~\ref{Activation}. 
Nevertheless, it should be pointed out the significant effect of the maximum of the stripping excitation functions around 8-12 MeV, shown in Fig.~\ref{FR_92Mo-100Mo}(b,e),  
on the increasing low--energy side and the maximum of the corresponding total activation excitation functions. 

To sum up, the deuteron total-reaction cross section that remains available for the PE+CN mechanisms has followed the due consideration of the incident flux leakage through the  breakup, stripping and pick-up DI processes. 
Using the present work notation as well as its first assessment of the deuteron break-up and DR transitions, the composite--formation cross section, e.g. Eq. (164) \cite{TALYS},  becomes $\sigma_{PE}$+$\sigma_{CN}$=$\sigma_R$-$\sigma_{BU}$-$\sigma_{DR}$. 
In brief, similarly to the depletion factor to account for the direct and PE effects within the default TALYS computation, Eq. (127) \cite{TALYS}, we have used finally a normalization factor of the deuteron total-reaction cross section, of the form:

\begin{eqnarray}\label{eq:2}
%
1 - \frac{\sigma_{BU} + \sigma_{DR}}{\sigma_R} \: 
\end{eqnarray}
for the final assessment of the PE+CN cross section. 
The energy dependence of this factor is shown in Fig.~\ref{FR_92Mo-100Mo}(c,f) for deuteron interaction with $^{92,100}$Mo target nuclei, at once with the fractions of the DI cross section and its BU and DR components to $\sigma_R$.  
One may note a steep increase with energy of PE+CN fraction, while the BU increase is much lower than that of ${\sigma_R}$. 
Next, PE+CN fraction reaches its maximum at deuteron energies of 15--20 MeV, and continues with a slow decrease due to the continuous BU increase with the incident energy. 

Actually, the weight of PE+CN versus DI fractions for an isotopic chain is triggered by the increase of the BU mechanism importance with increase of the target--nucleus mass number. 
Overall, the DI and statistical processes fractions vary around half of $\sigma_{R}$ \cite{Crd,Fed,Nid,Zrd} along the actual incident--energy range, pointing out the quite important DI role. 

\subsection{Statistical emission}
\label{PE+CN}

The statistical PE+CN mechanisms, which complete the deuteron interaction analysis along an enlarged nuclear-interaction time scale, become important with the increase of the incident energy above the Coulomb barrier. 
The corresponding reaction cross sections have been calculated using TALYS-1.96 code \cite{TALYS}, taking into account also the above--discussed BU approach as well as the stripping  and pick--up mechanisms through the normalization factor of the OMP total--reaction cross section. 
Another particular issue of the present calculations is the use of the same model parameters to account for different reaction mechanisms, as the same OMP parameters for calculation of the distorted waves in the ingoing/outgoing DR channels, the PE transition rates, and the transmission coefficients of various CN channels. 

The following input options of the TALYS code were used in the current work: 
(a) the OMPs of Koning-Delaroche \cite{KD}, Daehnick {\it et al.} \cite{dah}, Becchetti--Greenlees \cite{BG}, and Avrigeanu {\it et al.} \cite{avr2014} for neutrons and protons, deuterons, tritons, and $\alpha$-particles, respectively, 
(b) the model for breakup reactions 
including the BF enhancement \cite{BF_talys,breakup}, 
(c) the back-shifted Fermi gas formula for the nuclear level density (NLD), and 
(d) the PE exciton model with analytical transition rates with energy--dependent matrix element, and spin distribution based on particle--hole state densities. 
Furthermore, likewise to the analysis of $^{87}$Y$^{m,g}$ activation by deuterons incident on $^{nat}$Zr \cite{Zrd}, better account of the measured ground and isomeric state excitation functions of $^{91}$Nb, $^{93}$Mo, and $^{93,94,96,97}$Tc residual nuclei has been obtained within actual state-of-art TALYS calculations by amending the spin distribution cut-off parameter by a factor of 0.25. 

Next to the above--mentioned sharing of around half of the deuteron total--reaction cross section by the PE+CN mechanisms (Fig.~\ref{FR_92Mo-100Mo}), the weight of every reaction mechanism is concerned in the following analysis of the available data for deuteron--induced reactions on $^{92,94,95,96,97,98,100,nat}$Mo at energies up to 50 MeV.

\begin{figure*} 
\includegraphics[width=0.65\textwidth]{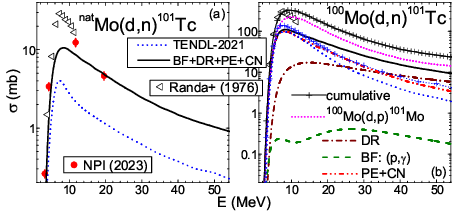}
\caption{(Color online) Comparison of NPI \cite{Mod} (solid circles) and previous \cite{randa76n} measurements, TENDL--2021 evaluation \cite{T21} (short dashed curves), and presently calculated (thick solid curves) cross sections of (a) $^{nat}$Mo$(d,n)^{101}$Tc, and (b) $^{100}$Mo$(d,n)^{101}$Tc$^{+}$ (see text), along with BF enhancement (dashed), $(d,n)$ 
stripping DR (dash-dotted), and PE+CN components (dash-dot-dotted), and contribution of $^{101}$Mo decay (dotted) to cumulative population of $^{101}$Tc (thin curves + cross)}
\label{101Tc_natMod}
\end{figure*}

\section{Results and Discussion} \label{Activation}

The excitation functions of residual nuclei from deuterons interaction with $^{nat}$Mo are compared in Figs.~\ref{101Tc_natMod}--\ref{89Zr-88Zr_natMod} with the available data corresponding also to irradiation of all Mo stable isotopes \cite{EXFOR}, the similar TENDL--2021 evaluation \cite{T21}, and the results of calculations using FRESCO \cite{FRESCO} and TALYS \cite{TALYS} codes. 
A particular attention is given, beyond the more recent data \cite{tak12,elbinawi,lebeda,tak11,chodash}, to new measured excitation functions at NPI CAS in \v{R}e\v{z} \cite{Mod} for the activation of $^{97}$Nb, $^{98}$Nb$^m$, $^{97}$Tc$^m$, and $^{101}$Tc residual nuclei.

The higher number of Mo stable isotopes, i.e. $^{92}$Mo (14.7\%), $^{94}$Mo (9.2\%),  $^{95}$Mo (15.9\%), $^{96}$Mo (16.7\%), $^{97}$Mo (9,6\%), $^{98}$Mo (24.2\%) and $^{100}$Mo (9.7\%),  made necessary the analysis of their detailed contributions to the activation of a certain residual nucleus. 
The involved reaction mechanisms are particularly illustrated in order to compare their strength. 
While some contributions are not shown for all Mo stable isotopes in Figs.~\ref{101Tc_natMod}--\ref{89Zr-88Zr_natMod}, they were all considered within the cross-section calculation for the natural Mo target. 

The activation of a certain residual nucleus is presented following four reaction sequences: 
$^{nat}$Mo$(d,xn)^{93-101}$Tc (Figs.~\ref{101Tc_natMod}--\ref{93Tc_natMod}), 
$^{nat}$Mo$(d,xnp)^{90-101}$Mo (Figs.~\ref{101Mo_natMod}--\ref{90Mo_natMod}), 
$^{nat}$Mo$(d,xn2p)^{90-98}$Nb (Figs.~\ref{98mNb-97Nb-96Nb_natMod}--\ref{92mNbP-91mNb-90Nb_natMod}), and 
$^{nat}$Mo$(d,xn3p)^{88,89}$Zr (Fig.~\ref{89Zr-88Zr_natMod}). 
The following discussion concerns firstly the heavier residual nuclei since fewer stable isotopes of Mo contribute to their population, while more isotopes,  processes, and emitted particles are involved going towards the lighter products.

\subsection{$^{nat}$Mo$(d,xn)^{93-101}$Tc reactions} \label{Activation-d,xn}

\subsubsection{$^{nat}$Mo$(d,n)^{101}$Tc}

The analysis of  $^{101}$Tc residual nucleus activation by deuterons incident on $^{nat}$Mo concerns only the deuteron interaction with the neutron-richest stable isotope $^{100}$Mo. 
Moreover, $^{101}$Tc excitation function has been for the first time reported \cite{Mod} for a natural molybdenum target [Fig.~\ref{101Tc_natMod}(a)], 
in addition to the previously measured $^{100}$Mo$(d,n)^{101}$Tc activation  cross sections \cite{randa76n} [Fig.~\ref{101Tc_natMod}(b)]. 

In fact, there are two ways of $^{101}$Tc residual nucleus population, i.e. via $(d,n)$ reaction and $\beta^{-}$ decay of $^{101}$Mo ($T_{1/2}$=14.61 min) which is activated through $^{nat}$Mo$(d,p)$ reaction (Sec.~\ref{Activation-d,p}). 
The NPI--reported $^{101}$Tc activation cross sections \cite{Mod} have been corrected for the $^{101}$Mo decay contribution, while the same correction was noticed also by Randa {\it et al.} within their reported data for the activation of $^{100}$Mo isotope. 
Moreover, amending Randa {\it et al.} data by $^{100}$Mo natural abundance, 
it results a good agreement of the two data sets below the incident energy of 10 MeV. 
However, the maximum of the wholly resulting excitation function in Fig.~\ref{101Tc_natMod}(a) is maybe followed by a too sharp decrease at the higher incident energies of the NPI measurement \cite{Mod}. 
 
The calculated cross sections for $^{100}$Mo$(d,n)^{101}$Tc reaction have proved a significant DR stripping $(d,n)$ contribution only for incident energies above 20 MeV, becoming larger than the statistical PE+CN from $\sim$30 MeV [Fig.~\ref{101Tc_natMod}(b)]. 
At the same time the inelastic breakup enhancement through $(p,\gamma)$ reaction remains much lower in the whole energy range. 
Overall, it results a notable underestimation of the measured cross--section maximum \cite{randa76n}, at the energies where PE+CN component is prevailing. 
A rather similar trend has the TENDL evaluation \cite{T21}, particularly at the energies where the DR weight is larger. 

On the other hand, the addition of the $^{101}$Mo decay has led for the cumulative calculated cross sections to an improved account of the  reported {$^{100}$Mo$(d,n)^{101}$Tc excitation--function maximum \cite{randa76n}. 
Its further decrease is however less quick, as it is that of the cumulative TENDL-2021 evaluation in Fig.~\ref{101Tc_natMod}(b), too.  

Nevertheless, the present approach has provided the rather good agreement of the newly measured $^{nat}$Mo$(d,n)^{101}$Tc cross sections shown in Fig.~\ref{101Tc_natMod}(a). 
This suitable description stands also as a proof of the appropriate correction for the removal of the  $^{101}$Mo decay contribution. 
The apparent underestimation of both data sets 
by the TENDL evaluation points out the effects of neglecting especially the stripping $(d,n)$ contribution which is larger than TENDL-2021 data above 15 MeV [Fig.~\ref{101Tc_natMod}(b)].

\begin{figure*} 
\includegraphics[width=0.9\textwidth]{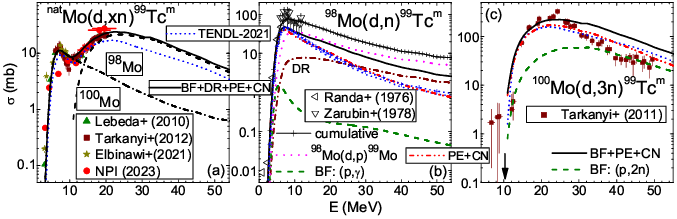}
\caption{(Color online) As Fig.~\ref{101Tc_natMod} but for $^{99}$Tc$^m$ activation \cite{tak12,elbinawi,EXFOR,lebeda,tak11,Mod,randa76n,zarubin} and 
(a) $^{nat}$Mo$(d,xn)^{99}$Tc$^m$, 
(b) $^{98}$Mo$(d,n)^{99}$Tc$^{m}$, and 
(c) $^{100}$Mo$(d,3n)^{99}$Tc$^m$ reaction cross sections, 
with (a) contributions of $(d,n)$ and $(d,3n)$ reactions (dash-dotted and dashed, respectively), 
and (c) an arrow at threshold energy (see text)} 
\label{99mTc_natMod}
\end{figure*}

\begin{figure*} 
\includegraphics[width=0.997\textwidth]{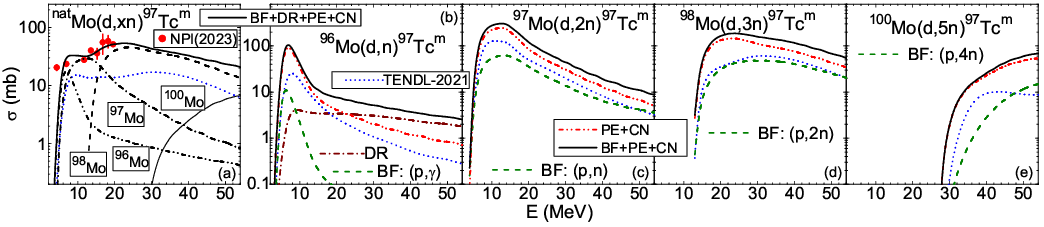}
\caption{(Color online) As Fig.~\ref{101Tc_natMod} but for $^{97}$Tc$^m$ activation \cite{Mod} and 
(a) $^{nat}$Mo$(d,xn)^{97}$Tc$^m$,  
(b) $^{96}$Mo$(d,n)^{97}$Tc$^m$, 
(c) $^{97}$Mo$(d,2n)^{97}$Tc$^{m}$, 
(d) $^{98}$Mo$(d,3n)^{97}$Tc$^{m}$, and 
(e) $^{100}$Mo$(d,5n)^{97}$Tc$^{m}$ reaction cross sections, 
with (a) contributions of $(d,n)$, $(d,2n)$, $(d,3n)$, and $(d,5n)$ reactions (dash-dot-dotted, dash-dotted, dashed, and thin solid curves, respectively)}
\label{97mTc_natMod}
\end{figure*}

\subsubsection{$^{nat}$Mo$(d,xn)^{99}$Tc$^m$}

The $^{nat}$Mo$(d,xn)^{99}$Tc$^m$ activation cross sections measured most recently at NPI \cite{Mod} are in the good agreement with the recent data \cite{tak12,elbinawi,lebeda} shown in Fig.~\ref{99mTc_natMod}(a). 
It should be noted again that the population of $^{99}$Tc$^m$ is also cumulative, including the $\beta^{-}$ decay of $^{99}$Mo ($T_{1/2}$=65.9 h) activated through $^{98}$Mo$(d,p)$ reaction (Sec.~\ref{Activation-d,xp}). 
The appropriate correction has been carried out within the NPI--measured values \cite{Mod}, as well as reported by Lebeda {\it et al.} \cite{lebeda}, Elbinawi {\it et al.} \cite{elbinawi}, and  T\'ark\'anyi {\it et al.} \cite{tak12} on both $^{nat}$Mo} and $^{100}$Mo \cite{tak11}. 
The same correction was noticed by Randa {\it et al.} \cite{randa76n} for $^{98}$Mo$(d,n)^{99}$Tc$^m$ reaction, but not by Zarubin {\it et al.} \cite{zarubin} for a similar measurement [Fig.~\ref{99mTc_natMod}(b)]. 

The calculated results of this work for the $^{nat}$Mo$(d,xn)^{99}$Tc$^m$ activation are in a rather good agreement with all data available until an incident energy of $\sim$20 MeV [Fig.~\ref{99mTc_natMod}(a)].  
More important within this energy range is firstly the $(d,n)$ reaction on $^{98}$Mo nucleus which, however, becomes half of order of magnitude lower than that of $(d,3n)$ reaction at 20 MeV. 
The statistical PE+CN component is the major one for the former reaction, below $\sim$30 MeV, while then this role is taken by the stripping DR [Fig.~\ref{99mTc_natMod}(b)]. 
It is not the same case of the latter reaction, where BF contribution increases significantly at higher energies but yet below the PE+CN one shown in Fig.~\ref{99mTc_natMod}(c). 

Thus, it has been of interest to understand a sizable underestimation of $^{98}$Mo$(d,n)^{99}$Tc$^m$ data of Zarubin {\it et al.} \cite{zarubin} at the same time with a suitable account of the similar data of Randa {\it et al.} \cite{randa76n} in Fig.~\ref{99mTc_natMod}(b). 
Addition of the above--mentioned decay of $^{99}$Mo residual nucleus, in a similar way to $^{100}$Mo$(d,p)^{101}$Tc reaction  analysis, has led anyhow to a cumulative excitation function in agreement with the former data, too. 

However, a suitable account of the measured $^{100}$Mo$(d,3n)^{99}$Tc$^m$ excitation function \cite{tak11} is found only until $\sim$30 MeV, the arrow in Fig.~\ref{99mTc_natMod}(c) pointing out the threshold of this reaction. 
Moreover, an overestimation is then apparent by even the statistical PE+CN component alone. 
The BF enhancement being already comparable with the experimental data, its strong  importance becomes obvious for the third--chance particle emission channel, in comparison to $(p,\gamma)$ BF added to $(d,n)$ one--particle emission. 
Overall, the model calculations lead to better data account versus TENDL evaluation while a particular extension of the measurements to incident energies over 30 MeV is highly requested. 

\subsubsection{$^{nat}$Mo$(d,xn)^{97}$Tc$^m$}

The measured $^{nat}$Mo$(d,xn)^{97}$Tc$^{m}$ excitation function has recently been reported for the first time \cite{Mod} while no measurement exists for $^{96,97,98,100}$Mo isotopes involved in this respect. 
Its account by the present calculations is related firstly to results for the $(d,n)$ and $(d,2n)$ reactions on $^{96}$Mo and $^{97}$Mo isotopes, respectively [Fig.~\ref{97mTc_natMod}(a)].
The $(d,3n)$ reaction channel for $^{98}$Mo target nucleus becomes also significant close to the incident energy of 20 MeV, while the $(d,5n)$ reaction on $^{100}$Mo is relevant for $^{97}$Tc$^{m}$ activation only above 50 MeV. 

A routine comment may concern the PE+CN major contribution to the population of $^{97}$Tc$^{m}$. 
However, the $(d,n)$ stripping component is shown in Fig.~\ref{97mTc_natMod}(b)  to be even more important at incident energies over $\sim$25 MeV. 
These DR results are, at their turn, exceeded at the low energies $\leq$9 MeV by the BF contribution through the $(p,\gamma)$ reaction even though followed by a fast decrease. 

On the other hand, a notable BF enhancement corresponds to the $(p,n)$, $(p,2n)$, and $(p,4n)$ reactions induced by the breakup--protons on the heavier isotopes $^{97,98,100}$Mo, respectively, with additional contributions to $(d,2n)$, $(d,3n)$, and $(d,5n)$ reactions on the same nuclei [Fig.~\ref{97mTc_natMod}(c-e)]. 
Nevertheless, the neglected additions of BF and especially DR processes might contribute to  the apparent underestimation of $^{nat}$Mo$(d,xn)^{97}$Tc$^m$ excitation function by TENDL-2021 in  Fig.~\ref{97mTc_natMod}(a).

\begin{figure*} 
\includegraphics[width=0.997\textwidth]{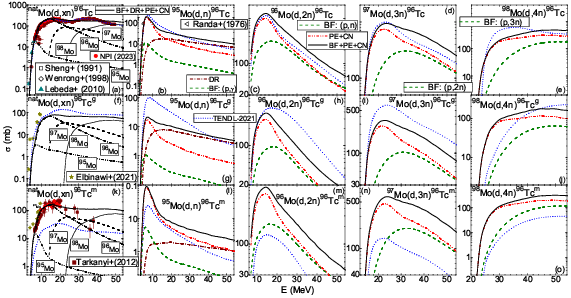}
\caption{(Color online) As Fig.~\ref{101Tc_natMod} but for 
(a-e) $^{96}$Tc, (f-j) $^{96}$Tc$^g$, and (k-o) $^{96}$Tc$^m$ activation \cite{tak12,elbinawi,EXFOR,lebeda,Mod,randa76n,wenrong,sheng},
with (a,f,k) contributions of $(d,n)$, $(d,2n)$, $(d,3n)$, and $(d,4n)$ reactions (dash-dot-dotted, dash-dotted, dashed, and thin solid curves, respectively)}
\label{96Tc_natMod}
\end{figure*}

\subsubsection{$^{nat}$Mo$(d,xn)^{96}$Tc$^{g+m,g,m}$}

The model calculations describe well the recent data \cite{Mod} as well as previous measurements of $^{96}$Tc$^{g+m,g,m}$ excitation functions \cite{tak12,elbinawi,EXFOR,lebeda,randa76n,wenrong,sheng} in Fig.~\ref{96Tc_natMod}(a,b,f,k). 
The contributions of specific outgoing channels and/or  Mo isotopes are shown in Fig.~\ref{96Tc_natMod}(a,f,k), the major role for $^{96}$Tc$^{g+m,g,m}$ activation changing from $(d,2n)$ reaction, for incident energies $\leq$24 MeV, to $(d,3n)$ and, finally, $(d,4n)$ reactions, with the incident energy increase.

Concerning the involved reaction mechanisms, PE+CN plays an important role  except the $(d,n)$ reaction [Fig.~\ref{96Tc_natMod}(b,g,l)]. 
In a similar way to the previous analysis of $^{97}$Tc$^m$ activation, there is a strong competition of the DR stripping and PE+CN mechanisms beyond, e.g. 10 MeV. 
The former becomes dominant at higher energies. 
The BF enhancement, due to the $(p,\gamma)$ reaction induced by breakup protons, also exceeds the stripping contribution at low incident energies, although it decreases faster.  

Moreover, there is a sound increase, by about two order of magnitudes, of the BF contributions through $(p,n)$ and $(p,2n)$ reactions [Fig.~\ref{96Tc_natMod}(c,d)]. 
They become comparable with the statistical PE+CN at energies around 50 MeV. 
The $(p,3n)$ BF contribution to the $(d,4n)$ forth--chance particle emission is strong, too [Fig.~\ref{96Tc_natMod}(e,j,o)], but yet lower than $(p,n)$ and $(p,2n)$ reactions. 
Nevertheless, the lack of data related to specific Mo isotopes beyond the only one set  corresponding to $^{95}$Mo$(d,n)^{96}$Tc reaction \cite{randa76n} restrains further comments on the  theoretical approach.

\begin{figure*} 
\includegraphics[width=0.997\textwidth]{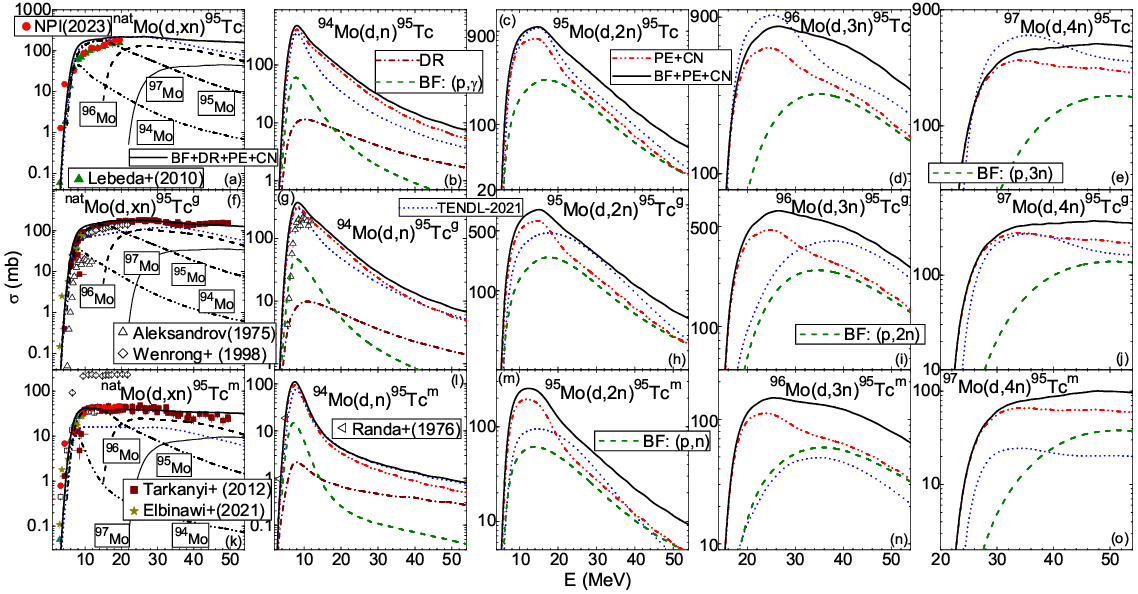}
\caption{(Color online)  As Fig.~\ref{96Tc_natMod} but for 
(a-e) $^{95}$Tc, (f-j) $^{95}$Tc$^g$, and (k-o) $^{95}$Tc$^m$ activation \cite{tak12,elbinawi,EXFOR,lebeda,Mod,randa76n,wenrong,sheng,alex75} (see text)}
\label{95Tc_natMod}
\end{figure*}

\subsubsection{$^{nat}$Mo$(d,xn)^{95}$Tc$^{g+m,g,m}$}

The $^{95}$Tc$^{g+m,m}$ activation cross sections include significant contributions of $^{94,95,96,97}$Mo isotopes, through reactions from $(d,n)$ to $(d,4n)$ respectively. 
The weight of the BU, DR, PE, and CN mechanisms, shown for each of these reactions in Fig.~\ref{95Tc_natMod}(b-e,g-j,l-o), may prove an appropriate account and validation of the present analysis for these reactions. 

However, similarly to the $^{96}$Tc$^{g+m,g,m}$ activation, very scarce related data of specific Mo isotopes exist only for $^{95}$Tc$^{g,m}$ populated through $(d,n)$ reaction \cite{randa76n,alex75}.
The present model analysis describes the measured excitation function for the latter state but overestimate the former [Fig.~\ref{95Tc_natMod}(g,l)]. 
At the same time, amending $^{94}$Mo$(d,n)^{95}$Tc$^g$ activation cross sections \cite{alex75} by natural abundance of $^{94}$Mo isotope, a discrepancy between them and $^{nat}$Mo$(d,xn)^{95}$Tc$^g$ data systematics becomes apparent in Fig.~\ref{95Tc_natMod}(f).

Again, the PE+CN dominant role within $^{95}$Tc$^{g+m,g,m}$ population at lower energies   is visible at incident energies above 20--30 MeV [Fig.~\ref{95Tc_natMod}(c-e,h-j,m-o)]. 
It is yet close to the inelastic breakup enhancement for $(d,xn)$ reactions, with $x$=2--4.
The competition between the inelastic breakup contribution, through the $(p,\gamma)$ reaction, and the DR stripping into the $(d,n)$ outgoing channel is also obvious in Fig.~\ref{95Tc_natMod}(b,g,l). 
The former exceeds the latter one at low incident energies, with an excitation function maximum higher by an order of magnitude at $\sim$8 MeV.
Then, a faster decrease is leading to the opposite case around 50 MeV. 
So, at the same energy, the DR contribution is however well below the PE+CN one. 

Once more, the need to enlarge the data systematics related to Mo stable isotopes activation, particularly at higher energies, should be pointed out. 

\begin{figure*} 
\includegraphics[width=0.9\textwidth]{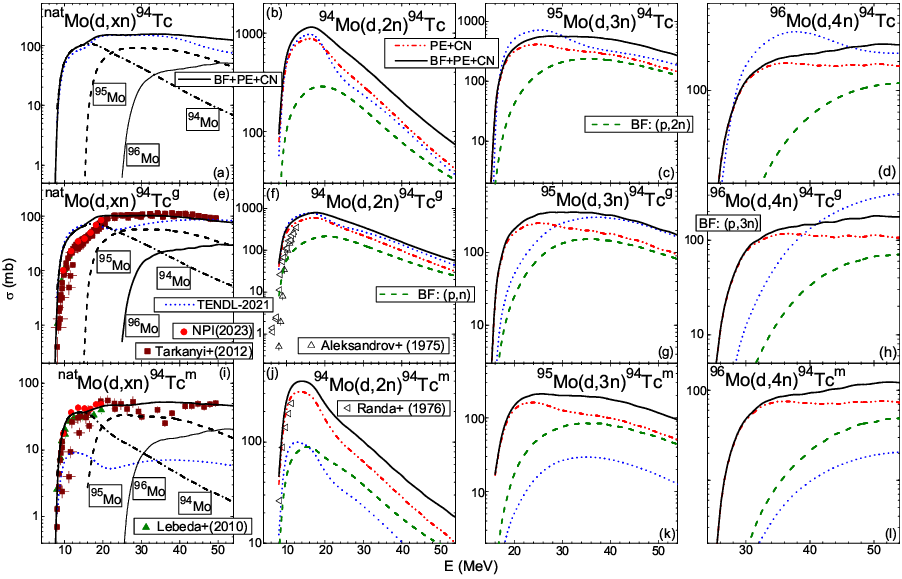}
\caption{(Color online) As Fig.~\ref{96Tc_natMod} but for 
(a-d) $^{94}$Tc, (e-h) $^{94}$Tc$^g$, and (i-l) $^{94}$Tc$^m$ activation \cite{tak12,EXFOR,lebeda,Mod,randa76n,alex75}, 
with (a,e,i) contributions of $(d,2n)$, $(d,3n)$, and $(d,4n)$ reactions (dash-dotted, dashed, and thin solid curves, respectively).
}
\label{94Tc_natMod}
\end{figure*}
 
\begin{figure*} 
\includegraphics[width=0.9\textwidth]{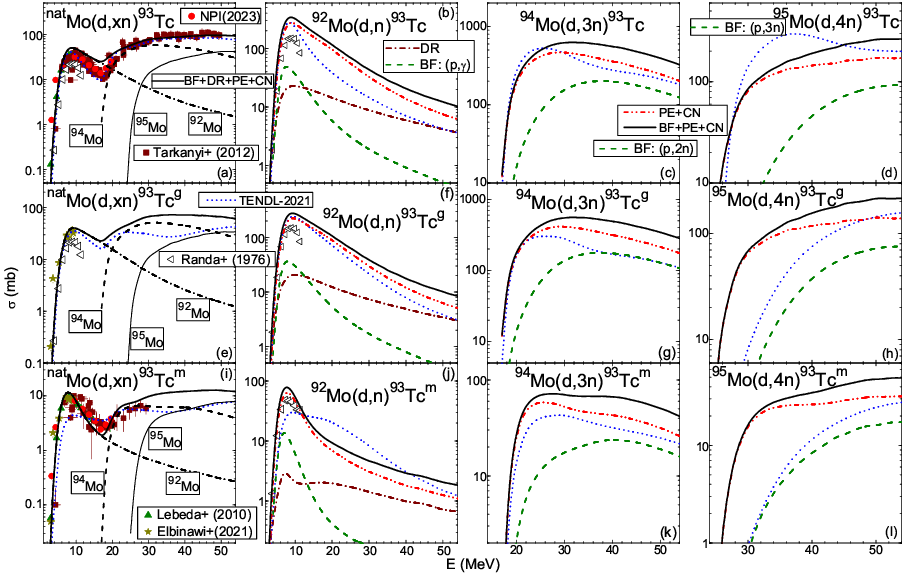}
\caption{(Color online)  As Fig.~\ref{94Tc_natMod} but for 
(a-d) $^{93}$Tc, (e-h) $^{93}$Tc$^g$, and (i-l) $^{93}$Tc$^m$ activation \cite{tak12,elbinawi,EXFOR,lebeda,Mod,randa76n}, 
with (a,e,i) contributions of $(d,n)$, $(d,3n)$, and $(d,4n)$ reactions (dash-dotted, dashed, and thin solid curves, respectively)}
\label{93Tc_natMod}
\end{figure*}

\subsubsection{$^{nat}$Mo$(d,xn)^{94}$Tc$^{g+m,g,m}$}

The actual model analysis shows a very good agreement with the measured excitation function for $^{94}$Tc$^{m}$ activation [Fig.~\ref{94Tc_natMod}(i)]. 
However, the $^{94}$Tc$^{g}$ activation data are well described just above the threshold as well as at incident energies above 20 MeV, while there is an overestimation between 10--20 MeV in Fig.~\ref{94Tc_natMod}(e). 

As already noticed above, scarce measurements for activation of specific Mo isotopes consist of two earlier data sets corresponding in Fig.~\ref{94Tc_natMod}(f,j) to $^{94}$Mo(d,2n)$^{94}$Tc$^{g,m}$ reactions \cite{randa76n,alex75}. 
Unfortunately, they were carried out only at the lower incident energies $\leq$12 MeV.  
The present model calculations describe the latter \cite{randa76n} but slightly overestimate the former \cite{alex75}. 

It is thus proved that similar overestimation of the $^{94}$Tc$^{g}$ activation is due to the $(d,2n)$ reaction on $^{94}$Mo, despite the lowest natural abundance of this isotope. 
One may further note the statistical nature of the related difference between experimental and calculated results, the only additional BF contribution through the $(p,n)$ reaction being lower by roughly one order of magnitude. 
Certainly, any further measurements of $^{94}$Tc activation on specific Mo isotopes as well as measured excitation functions of $^{nat}$Mo$(d,xn)^{94}$Tc$^{g+m,g}$ over 20 MeV incident energy would be most helpful for detailed analysis of these discrepancies.

\subsubsection{$^{nat}$Mo$(d,xn)^{93}$Tc$^{g+m,g,m}$}

The model calculations describe the excitation functions of $^{93}$Tc$^{g+m,g,m}$  activation by deuterons on natural Mo, as well as the data corresponding to $^{94}$Mo$(d,n)^{93}$Tc$^{m}$ reaction \cite{randa76n} in Fig.~\ref{93Tc_natMod}(j). 

There is however an overestimation of the measured $^{92}$Mo$(d,n)^{93}$Tc$^{g+m,g}$ excitation functions \cite{randa76n} in Fig.~\ref{93Tc_natMod}(b,f) below an incident energy of 12 MeV.  
Moreover, amending the activation cross sections of the $^{92}$Mo$(d,n)$ reaction \cite{randa76n} by the natural abundance of $^{92}$Mo, the corresponding derived data in Fig.~\ref{93Tc_natMod}(a,e) are lower, too, than existing systematics for natural Mo target. 

It should be noted also in this case the BF enhancement of the $(d,n)$, $(d,3n)$, and $(d,4n)$ cross sections through the $(p,\gamma)$, $(p,2n)$, and $(p,3n)$ reactions, respectively, induced by the breakup protons. 
However, its addition to the statistical PE+CN contributions in Fig.~\ref{93Tc_natMod}(b-d,f-h,j-l) is important only at incident energies higher than 20--30 MeV. 
At the same time, the sizable maximum of $(p,\gamma)$ excitation function exceeds the $(d,n)$ stripping DR at $\sim$10 MeV. 
Nevertheless, its further quick decrease makes the DR component to have the same role in the $(d,n)$ reaction [Fig.~\ref{93Tc_natMod}(b,f,j)] as the BF one in the $(d,xn)$ reactions, with x$\geq$2 [Fig.~\ref{93Tc_natMod}(c,d,g,h,k,l)]. 

\begin{figure*} 
\includegraphics[width=0.65\textwidth]{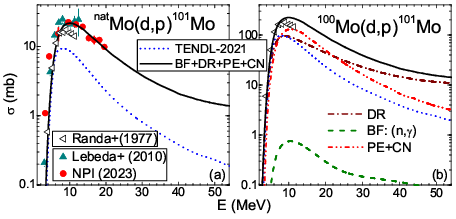}
\caption{(Color online) As Fig.~\ref{101Tc_natMod} but for $^{101}$Mo activation cross sections \cite{EXFOR,lebeda,Mod,randa77p} of (a) $^{nat}$Mo$(d,p)^{101}$Mo, and (b) $^{100}$Mo$(d,p)^{101}$Mo reactions (see text)}
\label{101Mo_natMod}
\end{figure*} 

\subsection{$^{nat}$Mo$(d,xnp)^{101-90}$Mo reactions} \label{Activation-d,xp}

\subsubsection{$^{nat}$Mo$(d,p)^{101}$Mo} \label{Activation-d,p}

Similarly to the case of $^{101}$Tc residual nucleus, the activation of $^{101}$Mo follows the deuteron interaction with only the neutron--richest $^{100}$Mo stable isotope. 
The model calculations, taking into account the BU, DR, and statistical PE+CN mechanisms, describe rather well the two experimental excitation functions \cite{lebeda,Mod} shown in Fig.~\ref{101Mo_natMod}(a). 
There are also measured activation cross sections corresponding to $^{100}$Mo$(d,p)^{101}$Mo reaction \cite{randa77p}, slightly overestimated in Fig.~\ref{101Mo_natMod}(b). 
Amending these cross sections by the natural abundance of $^{100}$Mo isotope, it was obtained a derived data set also lower than both newer measurements in the former figure. 

The stripping $(d,p)$ process provides now a much strong contribution to $^{101}$Mo activation comparing with $(d,n)$ one shown in Fig.~\ref{101Tc_natMod}(b) for $^{101}$Tc activation. 
Actually, the description of $^{nat}$Mo$(d,p)^{101}$Mo excitation function, with contributions only from a single  Mo isotope, is relevant for the importance of the stripping $(d,p)$ mechanism. 
This proof is just in line with the previous analyses of the experimental $(d,p)$ excitation functions which can not be described as long as the substantial DR stripping contribution is neglected \cite{VCod,Crd,Mnd,Fed,Nid,Zrd,Nbd}. 
Moreover, the apparent underestimation by TENDL evaluation of both $^{nat,100}$Mo$(d,p)^{101}$Mo reactions (Fig.~\ref{101Mo_natMod}) follows entirely the lack of taking into consideration the DR stripping process.

\subsubsection{$^{nat}$Mo$(d,xnp)^{99}$Mo}

\begin{figure*} 
\includegraphics[width=0.9\textwidth]{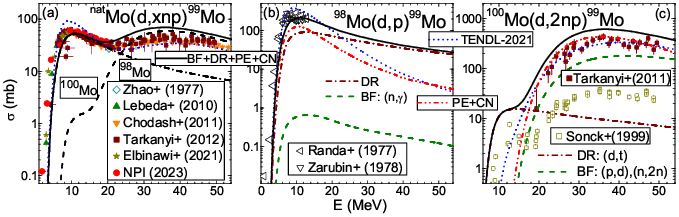}
\caption{(Color online)  
As Fig.~\ref{101Mo_natMod} but for $^{99}$Mo activation cross sections \cite{tak12,elbinawi,EXFOR,lebeda,tak11,chodash,Mod,zarubin,wenrong,randa77p,sonck} of  
(a) $^{nat}$Mo$(d,xnp)^{99}$Mo, 
(b) $^{98}$Mo$(d,p)^{99}$Mo, and 
(c) $^{100}$Mo$(d,2np)^{99}$Mo reaction cross sections, 
with (a) contributions of $(d,p)$ and $(d,2np)$ reactions (dash-dotted and dashed, respectively)} 
\label{99Mo_natMod}
\end{figure*}

The analysis of $^{99}$Mo activation by deuterons interaction with natural Mo and its stable isotopes $^{98,100}$Mo could take advantage of the existing systematics for Mo target \cite{tak12,elbinawi,EXFOR,lebeda,chodash,Mod,wenrong} shown in Fig.~\ref{99Mo_natMod}(a) as well as the measured excitation functions for $^{98}$Mo \cite{zarubin,randa77p} and $^{100}$Mo \cite{tak11,sonck} target nuclei, respectively [Fig.~\ref{99Mo_natMod}(b,c)]. 
In fact, it involves two direct reactions of stripping $^{98}$Mo$(d,p)^{99}$Mo and pick-up $^{100}$Mo$(d,t)^{99}$Mo, along with the breakup and PE+CN statistical emission. 
Their weight is however quite distinct for each of the two isotopes and even within different energy ranges, so that their specific discussion is required.

Thus, the DR stripping mechanism prevails over PE+CN in the case of $^{98}$Mo$(d,p)^{99}$Mo excitation function above the incident energy of 15 MeV, while at lower energies they are close to each other. 
The measured $^{98}$Mo$(d,p)^{99}$Mo data \cite{zarubin,randa77p} are thus described rather well [Fig.~\ref{99Mo_natMod}(b)], while the BF enhancement through breakup--neutron induced reaction $^{98}$Mo$(n,\gamma)^{99}$Mo is essentially negligible. 
Unfortunately, lack of data at energies above 15 MeV makes not possible a more detailed analysis.

Comparatively, the BF contribution through $(n,2n)$ and $(p,d)$ reactions is shown in Fig.~\ref{99Mo_natMod}(c) to be strong for $^{100}$Mo$(d,2np)^{99}$Mo reaction. 
It is crossing the pick-up $(d,t)$ excitation function at $\sim$20 MeV, becoming higher  by one order of magnitude at $\sim$30 MeV. 
However, it should be pointed out this pick-up reaction as the only contribution to $^{99}$Mo activation by deuterons on $^{100}$Mo at the incident energies $\leq$ 14 MeV. 
Nevertheless, it is by an order of magnitude lower than the $(d,p)$ stripping reaction on $^{98}$Mo, and thus not visible in the case of $^{99}$Mo activation by deuterons on the natural Mo target.

Finally, the sum of BF, DR, and PE+CN contributions overestimates the dispersed  $^{100}$Mo$(d,2np)^{99}$Mo activation data \cite{tak12,sonck}. 
Consequently, a similar overestimation exists above 30 MeV for the $^{nat}$Mo$(d,p)^{99}$Mo excitation function [Fig.~\ref{99Mo_natMod}(a)]. 
On the other hand, the suitable account of this excitation function up to $\sim$20 MeV does validate the $(d,p)$ stripping modeling. 
New measurement at higher energies may show if there is a major question of the actual analysis in this respect.

\begin{figure*} 
\includegraphics[width=0.99\textwidth]{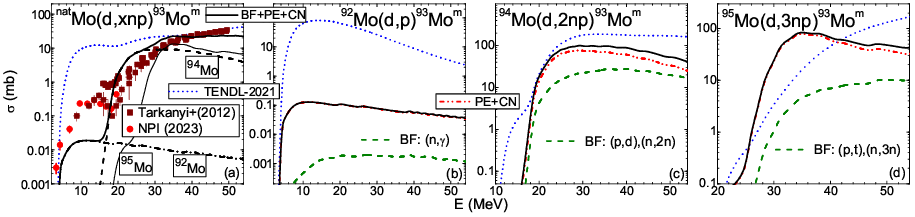}
\caption{(Color online) As Fig.~\ref{101Mo_natMod} but for $^{93}$Mo$^m$ activation \cite{tak12,Mod} and 
(a) $^{nat}$Mo$(d,xnp)^{93}$Mo$^m$, 
(b) $^{92}$Mo$(d,p)^{93}$Mo$^m$, 
(c) $^{94}$Mo$(d,2np)^{93}$Mo$^m$, and 
(d) $^{95}$Mo$(d,3np)^{93}$Mo$^m$ reaction cross sections, 
with (a) contributions of $(d,p)$, $(d,2np)$, and $(d,3np)$ reactions (dash-dotted,  dashed, and thin curves, respectively)} 
\label{93mMo_natMod}
\end{figure*}
\begin{figure*} 
\includegraphics[width=0.9\textwidth]{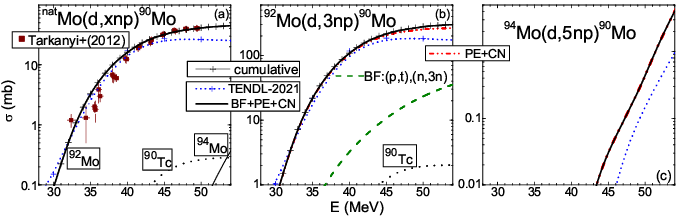}
\caption{(Color online) As Fig.~\ref{99Mo_natMod} but for $^{90}$Mo activation \cite{tak12}, 
with contributions of the $(d,3np)$ and $(d,5np)$ reactions (dashed and thin solid curves, respectively), as well as (a,b) $^{90}$Tc decay (dotted) to cumulative population (thin curves + cross) of $^{90}$Mo}  
\label{90Mo_natMod}
\end{figure*}

\subsubsection{$^{nat}$Mo$(d,xnp)^{93}$Mo$^m$}

The analysis for activation of the high--spin  $21{^+}$/2 isomeric state of $^{93}$Mo, at 2425.9 keV excitation energy, takes advantage of a close agreement [Fig.~\ref{93mMo_natMod}(a)] between the NPI recent measurement \cite{Mod} and the data obtained by T\'ark\'anyi {\it et al.} \cite{tak12} within their latest  irradiation in 2007, during fifteen years (1994--2007) studies at different accelerators.
In fact, a large spread between these results and the earlier ones (1995) \cite{tak12}  has been just at the energies of the latest measurement \cite{Mod}, while more approaching data were found at higher energies. 
Otherwise, a model analysis is restrained by the lack of any measurement of this activation within deuteron--induced reactions on either $^{92}$Mo, $^{94}$Mo, or $^{95}$Mo isotopes that may have major related yields. 

Furthermore, a complementary modeling should take into consideration 
the DR mechanism for the striping $^{92}$Mo$(d,p)$ as well as pick-up $^{94}$Mo$(d,t)$ reactions, in addition to also more complex BF enhancements and, finally, the statistical PE+CN emission. 
However, there is no measured angular distribution corresponding to this $^{93}$Mo isomeric state either in $(d,p)$ stripping, or in $(d,t)$ pick-up reactions, in order to extract the corresponding spectroscopic factors of the stripped/picked neutron needed for the calculations of DR corresponding yield. 
Moreover, the calculation results for the total stripping and pick-up cross sections, shown in Fig.~\ref{FR_92Mo-100Mo}(b), do not include excited states of $^{93}$Mo which may populate $^{93}$Mo$^m$ through their decay. 
Therefore, no DR component has been possible to be added to the BF+PE+CN calculated cross sections in Fig.~\ref{93mMo_natMod}(b,c). 
Consequently, an underestimated excitation function could be expected at low incident energies in the absence of both stripping and pick-up contributions. 

At the same time it should be noted that the model results shown in Fig.~\ref{93mMo_natMod} have been obtained using an additional factor of 0.25 for the NLD spin cut-off parameter of the residual nucleus $^{93}$Mo. 
Comparing with the analysis of $^{87}$Y$^{g,m}$ activated by deuteron interaction with $^{nat}$Zr target \cite{Zrd}, there is no measurement of $^{93}$Mo$^g$ excitation function to confirm this spin cut-off adjustment on the basis of the corresponding $g$/$m$ ratio. 
This adjustment could be the reason of the apparent differences between the present and TENDL-2021 predictions in Fig.~\ref{93mMo_natMod}, too.

Nevertheless, one may note that the large amount of measured isomeric cross sections for the $^{93}$Nb$(d,2n)^{93}$Mo$^m$ \cite{Nbd} and $^{94}$Mo$(n,2n)^{93}$Mo$^m$ \cite{inertia1} reactions was already well described using the same breakup approach and no spin cut-off adjustment but the particle--hole state density (PSD) \cite{ma98,ah98} within the geometry-dependent hybrid PE model \cite{mb83}. 
This PSD formula has included the PE spin distribution that was early discussed by Feshbach {\it et al.} \cite{hf80} and further detailed by Fu \cite{cyf86}. 
These results are in line with the more recent conclusion that reduced values of the spin cut-off parameter, obtained from isomeric cross--section analysis, were artificial and resulted from the improper use of the CN spin distribution also for the PE spin distribution \cite{tk06}.

\subsubsection{$^{nat}$Mo$(d,xnp)^{90}$Mo}

The present analysis describes the data of T\'ark\'anyi {\it et al.} \cite{tak12} in Fig.~\ref{90Mo_natMod}(a). 
While only the $^{92}$Mo target nucleus [Fig.~\ref{90Mo_natMod}(b)] plays a role in this respect, versus $^{94}$Mo isotope [Fig.~\ref{90Mo_natMod}(c)], a contribution due to the decay of $^{90}$Tc residual nucleus ($T_{1/2}$=50.7 s) may also be noted. 
It follows the $(d,4n)$ reaction on $^{92}$Mo, being however lower by at least two orders of magnitude than $^{90}$Mo activation data and the corresponding cumulative excitation function. 

Moreover, there are no measurements of the $^{90}$Mo activation by deuterons incident on the distinct $^{92,94}$Mo isotopes. 
Thus, the analysis corresponding to the natural Mo is the only way for validation of the contributions from BF and PE+CN mechanisms, the latter being shown in Fig.~\ref{90Mo_natMod}(b,c) to be obviously the dominant one. 

The TENDL-2021 predictions in Fig.~\ref{90Mo_natMod}(a,b) are also cumulative, including the $^{90}$Tc decay. 
However, the underestimation by the TENDL-evaluation at higher energies may be just due to the yet neglected contribution of the inelastic breakup enhancement.

\subsection{$^{nat}$Mo$(d,xn2p)^{98-89}$Nb reactions}

\begin{figure*} [!htbp]
\includegraphics[width=0.997\textwidth]{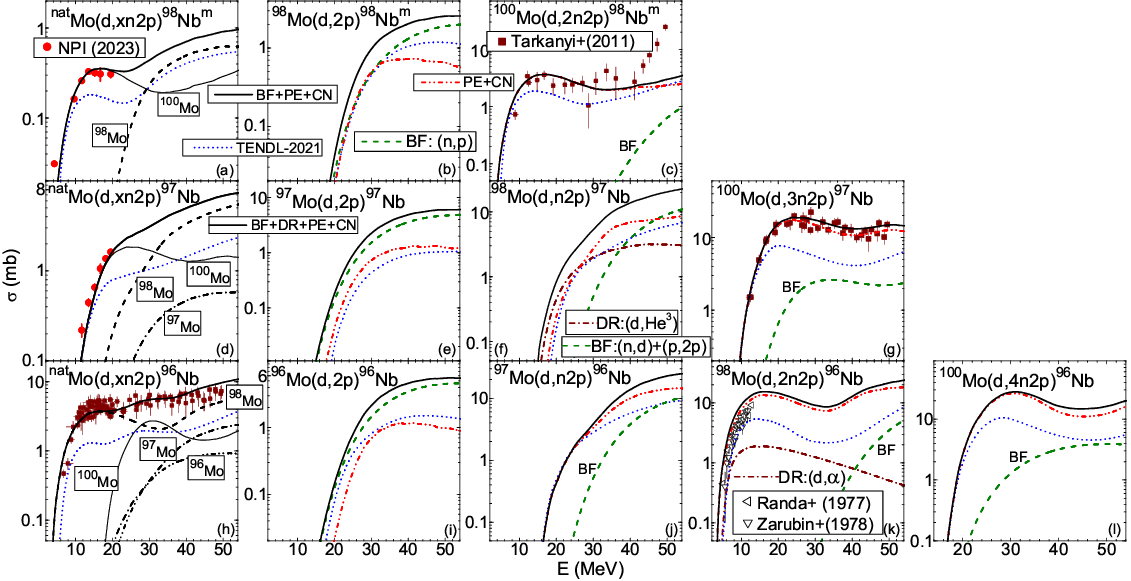}
\caption{(Color online) As Fig.~\ref{93mMo_natMod} but for 
(a)-(c) $^{98}$Nb$^{m}$, 
(d-g) $^{97}$Nb, and 
(h-l) $^{96}$Nb activation \cite{tak11,Mod,zarubin,randa77a}, with contributions of 
(a) the $(d,2p)$ and $(d,2n2p)$, 
(d) the $(d,2p)$, $(d,n2p)$, and $(d,3n2p)$, 
(h) the $(d,2p)$, $(d,n2p)$, $(d,2n2p)$, and $(d,4n2p)$ reactions (dash-dot-dotted, dash-dotted, dashed, and thin solid curves, respectively). }
\label{98mNb-97Nb-96Nb_natMod}
\end{figure*} 

\subsubsection{$^{nat}$Mo$(d,xn2p)^{98}$Nb$^m$}

A rather equally--well account of the excitation functions measured either recently \cite{Mod} for $^{nat}$Mo$(d,xn2p)^{98}$Nb$^m$  reaction [Fig.~\ref{98mNb-97Nb-96Nb_natMod}(a)] or formerly for  $^{100}$Mo$(d,2n2p)^{98}$Nb$^m$ reaction by T\'ark\'anyi {\it et al.} \cite{tak11} [Fig.~\ref{98mNb-97Nb-96Nb_natMod}(c)] has been shown by the actual analysis. 
Actually, only the two heaviest Mo isotopes may contribute to the activation of this isomeric state, through $^{98}$Mo$(d,2p)^{98}$Nb$^m$ and $^{100}$Mo$(d,2n2p)^{98}$Nb$^m$ reactions. 
The latter reaction channel corresponds, especially at the energies lower than the $(d,2n2p)$ reaction threshold, also to the $(d,\alpha)$ reaction with a maximum around the incident energy of 16 MeV. 

Because of the significantly higher effective threshold of the $(d,2p)$ reaction on $^{98}$Mo [Fig.~\ref{98mNb-97Nb-96Nb_natMod}(b)], the $^{98}$Nb$^m$ activation for the natural Mo target comes fully from this $(d,\alpha)$ reaction.
The obvious correspondence between the two above--mentioned data sets is emphasized within the actual BF+PE+CN analysis except the less usual steep increase of the latter excitation function at incident energies over $\sim$45 MeV. 
However, this model analysis has been proved difficult due to the lack of any spectroscopic information concerning the pick-up $^{100}$Mo$(d,\alpha)^{98}$Nb$^m$ reaction. 
Therefore, beyond the dominant PE+CN statistical mechanisms, only the inelastic breakup has been involved in this particular case. 
Its correspondence to a breakup--neutron induced reaction $(n,2np)$ led to cross--section contribution at incident energies only above 35 MeV and too low to account for the above--mentioned increase in Fig.~\ref{98mNb-97Nb-96Nb_natMod}(c). 

On the other hand, the comparison of the $(d,2p)$ and $(d,2n2p)$ interaction processes points out the  dominant inelastic breakup enhancement brought for the former reaction by breakup neutrons through $^{98}$Mo$(n,p)^{98}$Nb$^m$ reaction [Fig.~\ref{98mNb-97Nb-96Nb_natMod}(b)]. 
It exceeds by three times the PE+CN contribution for the $(d,2p)$ reaction on same target nucleus. 
This proton--emission enhancement is essential for the analysis of deuteron radiation--damage estimation in the structural materials, due to the additional hydrogen accumulation known as the "gas bubble accumulation". 

\subsubsection{$^{nat}$Mo$(d,xn2p)^{97}$Nb}

Similarly to the above, there is a suitable account of $^{97}$Nb activation data first--time recently reported \cite{Mod} for deuterons incident on natural Mo [Fig.~\ref{98mNb-97Nb-96Nb_natMod}(d)] as well as of the earlier measurement \cite{tak11} of the $(d,3n2p)$ reaction on $^{100}$Mo [Fig.~\ref{98mNb-97Nb-96Nb_natMod}(g)]. 
There is also a correspondence in the last case with a first excitation--function maximum around the incident energy of 23 MeV due to a $(d,n\alpha)$ reaction component with an obviously related low--energy side of the data for the natural Mo.

There is also a straightforward strong increase of the BF enhancement, being by far the dominant mechanism for $^{97}$Mo$(d,2p)^{97}$Nb reaction [Fig.~\ref{98mNb-97Nb-96Nb_natMod}(e)]. 
It overcomes the PE+CN contribution in $^{98}$Mo$(d,n2p)^{97}$Nb process too, for the incident energies $\geq$45 MeV shown in Fig.~\ref{98mNb-97Nb-96Nb_natMod}(f).

In fact, the availability of spectroscopic data related to the $^{98}$Mo$(d,^3He)^{97}$Nb pick-up reaction has been followed by a DR cross--section assessment. 
It has been thus proved the dominant role of this mechanism at incident energies $\leq$25 MeV, while it is still higher than the BF component up to $\sim$37 MeV [Fig.~\ref{98mNb-97Nb-96Nb_natMod}(f)].

\subsubsection{$^{nat}$Mo$(d,xn2p)^{96}$Nb}

The analysis of the $^{96}$Nb activation is of prime interest for the present analysis due to measured excitation function by T\'ark\'anyi {\it et al.} \cite{tak11}, with the lower incident energies $\leq$20 MeV corresponding fully to the  $(d,\alpha)$ reaction on the most abundant isotope  $^{98}$Mo [Fig.~\ref{98mNb-97Nb-96Nb_natMod}(h,k)]. 
There are also concerned contributions from $^{96,97,98,100}$Mo isotopes \cite{zarubin,randa77a}, through various BF, DR, PE, and CN mechanisms [Fig.~\ref{98mNb-97Nb-96Nb_natMod}(h-l)]. 

The model calculations describe much better the more recently measured $^{nat}$Mo$(d,xn2p)^{96}$Nb data \cite{tak11} than the earlier $^{98}$Mo$(d,2n2p)^{96}$Nb excitation function \cite{zarubin,randa77a} at the energies of the only $(d,\alpha)$ reaction. 
There are thus validated the major CN+PE statistical emission as well as the smaller but yet notable DR component in Fig.~\ref{98mNb-97Nb-96Nb_natMod}(k). 

Nevertheless, a main point is still the BF mechanism, with the $(n,p)$ enhancement bringing the dominant contribution to the $^{96}$Mo$(d,2p)^{96}$Nb activation. 
It completes thus the $^{98,97}$Nb activation in Fig.~\ref{98mNb-97Nb-96Nb_natMod}(b,e,i). 
The basic role of inelastic breakup and pick-up reaction contributions is confirmed also by the data underestimation of TENDL-2021 predictions \cite{T21} which are yet not including the account of these processes.

\begin{figure*}[!htbp]
\includegraphics[width=0.997\textwidth]{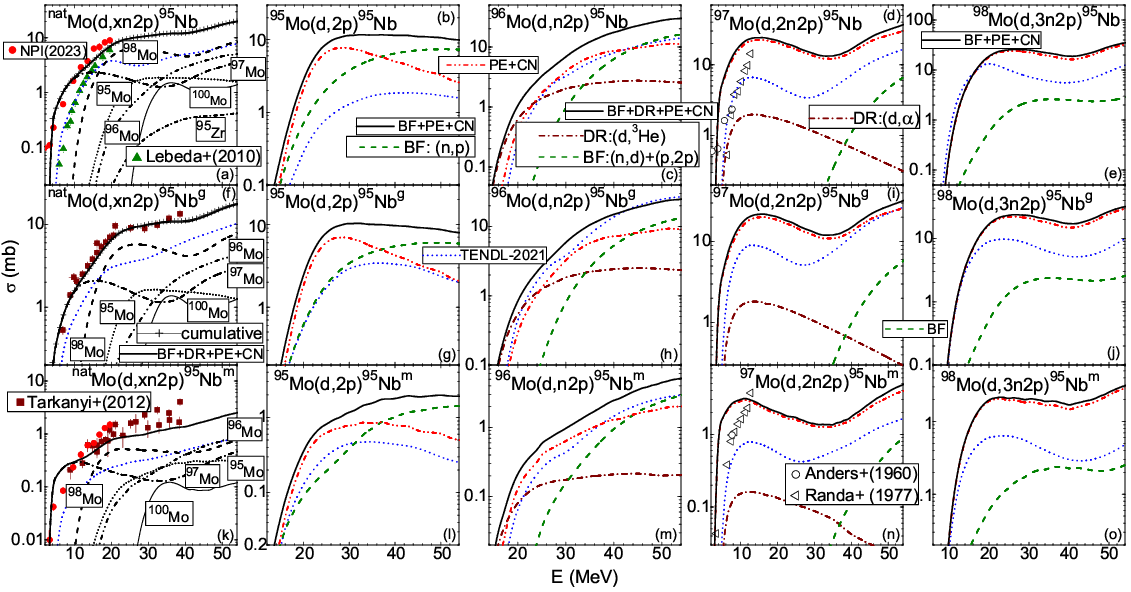}
\caption{(Color online) As Fig.~\ref{90Mo_natMod} but for 
(a-e) $^{95}$Nb, (f-j) $^{95}$Nb$^g$, and (k-o) $^{95}$Nb$^m$ activation \cite{tak12,EXFOR,lebeda,Mod,randa77a,anders},
with (a,f,k) contributions of $(d,2p)$, $(d,n2p)$, $(d,2n2p)$, $(d,3n2p)$, and $(d,5n2p)$ reactions (short-dotted, dash-dot-dotted, dash-dotted, dashed, and thin solid curves, respectively), and (a,f) contribution of the $^{95}$Zr decay (short-dash-dotted) to cumulative population (thin curves + cross) of $^{95}$Nb and $^{95}$Nb$^g$, respectively}
\label{95NbP_natMod}
\end{figure*} 

\subsubsection{$^{nat}$Mo$(d,xn2p)^{95}$Nb}

The actual BF+DR+PE+CN model results for $^{95}$Nb and $^{95}$Nb$^m$ activation, except for the points around the deuteron energy of $\sim$7 MeV in Fig.~\ref{95NbP_natMod}(a,k), are validated by recently measured cross sections through the $(d,xn2p)$ reactions on natural Mo \cite{Mod}, with well increased accuracy vs previous data \cite{tak12,lebeda}.

The same is true for the data reported more recently by T\'ark\'anyi {\it et al.} \cite{tak12}, including an additional excitation function of $^{nat}$Mo$(d,xn2p)^{95}$Nb$^g$ activation [Fig.~\ref{95NbP_natMod}(f)]. 
A weak contribution of $^{95}$Zr decay ($T_{1/2}$=65.9 h), visible only in Fig.~\ref{95NbP_natMod}(a), has been considered too, with no real difference for the $^{95}$Nb$^{g+m,g}$ cumulative activation. 
There are three points of this analysis which demands additional comments.

First, the most abundant isotope $^{98}$Mo brings the dominant contribution to $^{95}$Nb$^{g+m,g,m}$ activation by $(d,3n2p)$ reaction on natural Mo above the deuteron energy of $\sim$16 MeV. 
A first maximum of the $(d,2n2p)$ reaction, at lower energies, corresponds in fact to the one--particle emission $(d,\alpha)$ reaction on the nearly least abundant isotope $^{97}$Mo [Fig.~\ref{95NbP_natMod}(d,i,n)].  
Then, the two--particle emission $(d,n\alpha)$ sort of $(d,3n2p)$ reaction on $^{98}$Mo has an excitation--function maximum at 22-25 MeV [Fig.~\ref{95NbP_natMod}(e,j,o)]. 
At the same time there are earlier data sets \cite{randa77a,anders} of $(d,\alpha)$ reaction on $^{97}$Mo which are largely overestimated. 
However, it should be noted that the closely related recent data for natural Mo are well accounted for. 

Second, it could be again pointed out the important contributions of BF enhancement to $(d,2p)$ reaction cross sections, for the $^{95}$Mo target nucleus [Fig.~\ref{95NbP_natMod}(b,j,l)], similarly to the above--discussed $^{96,97,98}$Nb activation. 
Thus, the BF component exceeds the PE+CN statistical emission at the deuteron energies above 35 MeV. 
The latter becomes however dominant with the increase of additional neutron emission in $(d,xn2p)$ reactions with $x$$\geq$2. 
This is true also for deuterons on $^{100}$Mo, with results shown only in Fig.~\ref{95NbP_natMod}(a,f,k) but similar mechanism contributions as for $^{98}$Mo isotope.

Third, cross--section increases due to the $(d,^3He)$ and $(d,\alpha)$ pick-up DRs are noteworthy particularly for the former. 
This one is even larger than CN+PE component, at the lower energies $\leq$13 MeV, and BF  below 30--34 MeV  [Fig.~\ref{95NbP_natMod}(c,h,m)]. 
However, the TENDL-2021 evaluation \cite{T21} has followed the low--energy side of the earlier measured excitation function \cite{lebeda,randa77a,anders} in Fig.~\ref{95NbP_natMod}(a,d,n). 
Nevertheless, its underestimation of the $^{nat}$Mo$(d,xn2p)^{95}$Nb$^{g,m}$ reaction cross sections could be related to the missing of BF and DR pick-up  contributions.

\begin{figure*}[!htbp]
\includegraphics[width=0.997\textwidth]{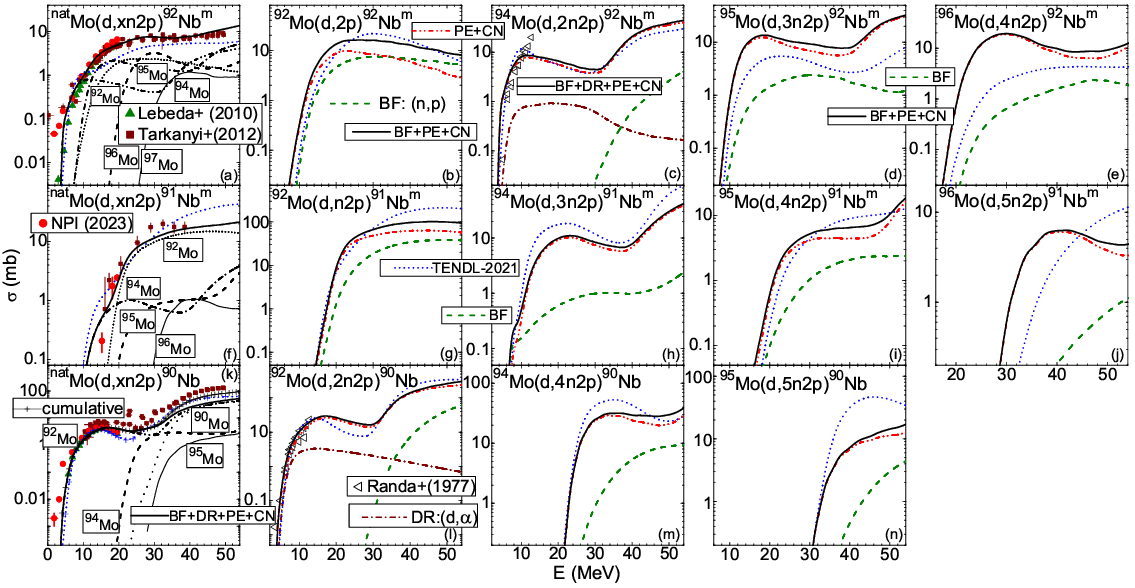}
\caption{(Color online) As Fig.~\ref{95NbP_natMod} but for 
(a-e) $^{92}$Nb$^{m}$, (f-j) $^{91}$Nb$^{m}$, and (k-n) $^{90}$Nb activation \cite{tak12,EXFOR,lebeda,Mod,randa77a}, 
with (a,f,k) contributions of $(d,2p)$, $(d,n2p)$, $(d,2n2p)$, $(d,3n2p)$, $(d,4n2p)$, and $(d,5n2p)$ reactions (short-dotted, short-dotted, dash-dot-dotted, dash-dotted, dashed, and thin-solid curves, respectively), and (k) $^{90}$Mo decay (short-dash-dotted)}
\label{92mNbP-91mNb-90Nb_natMod}
\end{figure*} 

\subsubsection{$^{nat}$Mo$(d,xn2p)^{92}$Nb$^m$}

The $^{nat}$Mo$(d,xn2p)^{92}$Nb$^m$ activation cross sections, measured recently at NPI with higher accuracy \cite{Mod}, as well as previous similar data \cite{tak12,lebeda} are rather well described by the actual model analysis except the lowest deuteron energies below 4 MeV [Fig.~\ref{92mNbP-91mNb-90Nb_natMod}(a)]. 
A similar agreement exists for an earlier measurement \cite{randa77a} at lower energies of the $(d,\alpha)$ reaction in Fig.~\ref{92mNbP-91mNb-90Nb_natMod}(c), as the first maximum of the $^{94}$Mo$(d,2n2p)$ excitation function, apart from data at $\sim$12 MeV. 
Nevertheless, there is a valuable account of the whole $^{nat}$Mo systematics at the same energy  [Fig.~\ref{92mNbP-91mNb-90Nb_natMod}(a)], with the largest contribution of $^{94}$Mo isotope in spite of its smallest natural abundance. 
The next important and main contribution up to the deuteron energy of 20 MeV, belongs to  $^{95}$Mo isotope. 
It corresponds also to a first maximum of the two--particle $(d,n\alpha)$ side of the $(d,3n2p)$ reaction channel [Fig.~\ref{92mNbP-91mNb-90Nb_natMod}(d)]. 

At the same time, the isotope $^{92}$Mo has the main role within $^{92}$Nb$^m$ activation  above 20 MeV, through the $(d,2p)$ reaction. 
The corresponding cross sections are again much increased by the BF enhancement due to the $^{92}$Mo$(n,p)$ reaction induced by breakup neutrons [Fig.~\ref{92mNbP-91mNb-90Nb_natMod}(b)]. 
This BF component becomes even higher than the CN+PE statistical processes at incident energies above 35 MeV. 
It is however overridden, up to the same energy, by the pick-up $(d,\alpha)$ reaction on $^{94}$Mo nucleus [Fig.~\ref{92mNbP-91mNb-90Nb_natMod}(c)].

\subsubsection{$^{nat}$Mo$(d,xn2p)^{91}$Nb$^m$}

Despite an increased uncertainty above its effective threshold, the available data of $^{91}$Nb$^m$ activation \cite{tak12,Mod} are rather well described taking into account the contributions of four isotopes and three reaction mechanisms [Fig.~\ref{92mNbP-91mNb-90Nb_natMod}(f-j)]. 
The dominant one in the whole energy range comes from the semi--magic nucleus $^{92}$Mo due to a BF enhancement close to main CN+PE statistical emission in Fig.~\ref{92mNbP-91mNb-90Nb_natMod}(g).  
It is exceeded only at energies below 20 MeV by the first maximum of the $(d,3n2p)$ excitation function 
for the target nucleus $^{94}$Mo, corresponding to the sequential emission of only a neutron and an $\alpha$-particle through actually a $(d,n\alpha)$ reaction (at variance with the sequential five--nucleon emission). 

Unfortunately, there is no measurement of this activation on any of the  Mo stable isotopes.

\begin{figure*} [!htbp]
\includegraphics[width=0.9\textwidth]{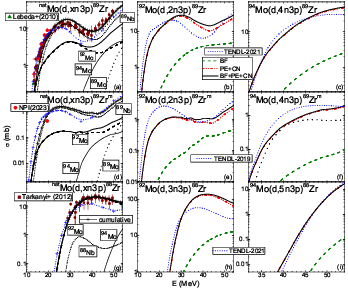}
\caption{(Color online) As Fig.~\ref{92mNbP-91mNb-90Nb_natMod}(k-m) but for 
(a-c) $^{89}$Zr, (d-f) $^{89}$Zr$^m$, and (g-i) $^{88}$Zr activation \cite{tak12,EXFOR,lebeda,Mod}, 
with contributions of (a,d,g) the $^{94,92}$Mo target isotopes (dashed and thin-solid curves, respectively), as well as (a,d) the $^{89}$Mo and $^{89}$Nb decay (dotted and short-dotted curves respectively), and (g) the $^{88}$Nb decay (short-dotted curve)}
\label{89Zr-88Zr_natMod}
\end{figure*}

\subsubsection{$^{nat}$Mo$(d,xn2p)^{90}$Nb}

Overall, the results of this analysis for $^{90}$Nb activation are in rather good agreement with the data measured for natural Mo and $^{92}$Mo targets \cite{tak12,lebeda,Mod,randa77a}. 
A note should concern the main contribution which is by far given by the two maxima of $^{92}$Mo$(d,2n2p)^{90}$Nb  excitation--function [Fig.~\ref{92mNbP-91mNb-90Nb_natMod}(l)]. 
The former, corresponding to the $\alpha$ emission channel, is also additionally increased by the $(d,\alpha)$ pick-up DR, while the BF has the same effect for the latter one, following the sequential nucleons evaporation. 
Actually, the CN+PE statistical emission mainly contributes to $^{90}$Nb activation while the breakup and pick-up processes have also a significant addition of $\sim$15\%.

The calculated results for $^{90}$Nb activation above 28 MeV had to include a steady contribution due to the decay of $^{90}$Tc ($T_{1/2}$=50.7 s) and consequently $^{90}$Mo ($T_{1/2}$=5.56 h), leading finally to a calculated cumulative $^{90}$Nb excitation function. 
This addition, involved similarly for the comparison with the TENDL-2021 evaluation, has been mostly useful to account for the data at deuteron energies above 20 MeV.

\begin{figure*}
\includegraphics[width=0.997\textwidth]{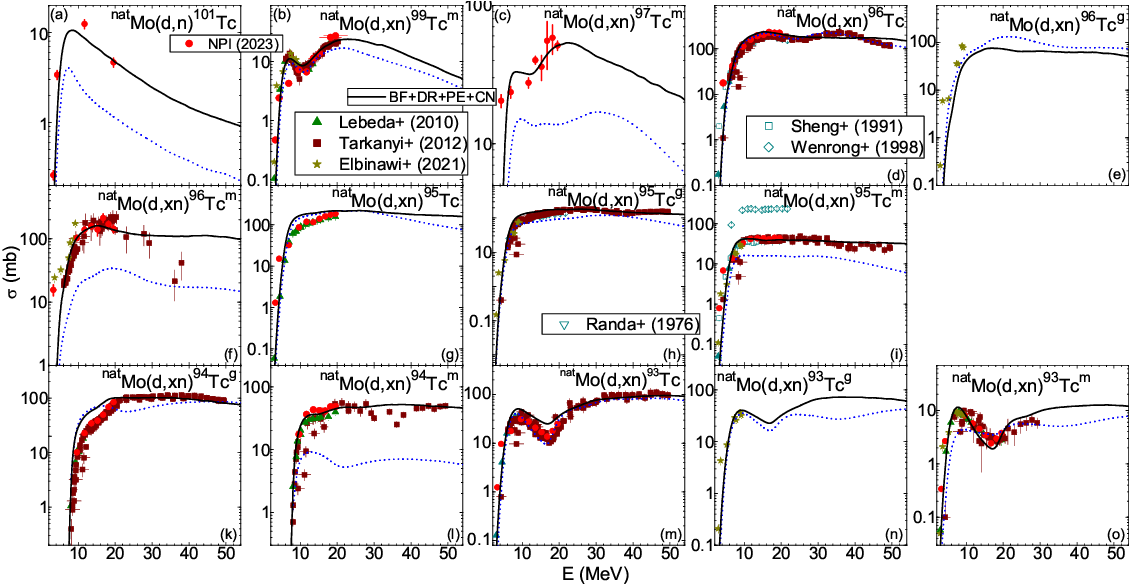}
\caption{(Color online)  Comparison of measured \cite{tak12,elbinawi,EXFOR,lebeda,tak11,Mod,randa76n,zarubin,wenrong,sheng,alex75}, TENDL-2021 \cite{T21} evaluated (dotted curves), and calculated (solid curves) cross sections for $^{nat}$Mo$(d,xn)^{93-101}$Tc excitation functions}
\label{natMod_d,xn}
\end{figure*}

\subsection{$^{nat}$Mo$(d,xn3p)^{89,88}$Zr reactions}

\subsubsection{$^{nat}$Mo$(d,xn3p)^{89}$Zr}

The measured  cross sections \cite{tak12,lebeda,Mod} for the activation of the residual nucleus $^{89}$Zr [Fig.~\ref{89Zr-88Zr_natMod}(a)] and $^{89}$Zr$^m$ state [Fig.~\ref{89Zr-88Zr_natMod}(d)] are cumulative, i.e. they include the decays of $^{89}$Tc ($T_{1/2}$=12.9 s) and $^{89}$Mo ($T_{1/2}$=2.11 min), and $^{89}$Nb ($T_{1/2}$=2.03 h) \cite{BNL}. 
Consequently, the present model analysis has taken into account, in addition to the $^{92,94}$Mo  isotopes contributions, these decays to $^{89}$Zr residual nucleus of $^{89}$Mo, including the $^{89}$Tc decay, as well as the decay of $^{89}$Nb. 
Thus, the experimental data are rather well described in Fig.~\ref{89Zr-88Zr_natMod}(a,d) while it should be pointed out the strongest contributions of $^{89}$Nb residual--nucleus decay versus the deuteron--induced reactions on $^{92,94}$Mo isotopes. 
The same addition has been considered for the comparison with the TENDL-2021 evaluations. 

Concerning the reaction mechanisms contributing to $^{89}$Zr residual--nucleus activation, the dominant PE+CN statistical emission is additionally increased at higher energies by the BF enhancement [Fig.~\ref{89Zr-88Zr_natMod}(b,c)].  
Particularly, this BF contribution is relevant to $^{92}$Mo$(d,2n3p)^{89}$Zr reaction. 
Actually, the first maximum of the $^{89}$Zr excitation function for both $^{92,nat}$Mo targets corresponds to $\alpha$-particle emission within $(d,p\alpha)$ reaction (Q=1.491 MeV), so important for the gas accumulation assessment.

\subsubsection{$^{nat}$Mo$(d,xn3p)^{88}$Zr}

The model analysis for the activation of $^{88}$Zr residual nucleus involves fewer issues and thus may have a more reliable outcome. 
Thus, it concerns contributions due to deuterons interaction mainly with $^{92,94}$Mo isotopes, as well as following $^{88}$Nb decay ($T_{1/2}$=7.78 min) \cite{BNL} taken into account for the present analysis results as well as well as in the case of the TENDL-2021 evaluations. 
A further contribution of $^{88}$Mo decay \cite{BNL} to $^{88}$Zr activation cross sections is not significant in the energy range of this work. 

The modeling results describe satisfactory the measured cumulative $^{88}$Zr excitation function \cite{tak12} shown in Fig.~\ref{89Zr-88Zr_natMod}(g). 
The largest contribution is related to the PE+CN mechanisms [Fig.~\ref{89Zr-88Zr_natMod}(h,i)], and also to the $^{92}$Mo isotope. 
An yet visible addition is due to the inelastic breakup enhancement in the case of $^{92}$Mo$(d,3n3p)$ reaction. 
The so weak contribution of $^{88}$Nb decay to $^{88}$Zr activation in Fig.~\ref{89Zr-88Zr_natMod}(g), in addition to $^{92,94}$Mo isotopes components, has pointed out the statistical emission model validation on this experimental basis.

\begin{figure*}
\includegraphics[width=0.997\textwidth]{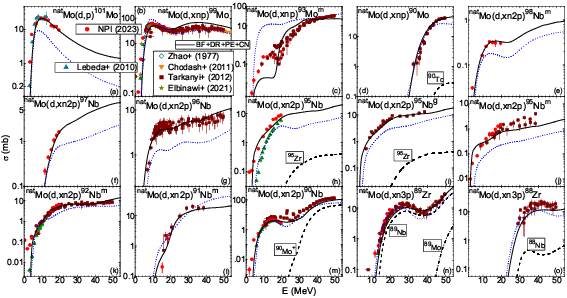}
\caption{(Color online) As Fig.~\ref{natMod_d,xn} but for $^{nat}$Mo$(d,xnp)^{90-101}$Mo, $^{nat}$Mo$(d,xn2p)^{90-98m}$Nb, and $^{nat}$Mo$(d,xn3p)^{88-89}$Zr reactions, with (d,h,i,m-o) $^{90}$Mo, $^{95}$Nb$^{g+m,g}$, $^{90}$Nb, $^{88}$Zr, and $^{89}$Zr cumulative activation including the decay of $^{90}$Tc, $^{95}$Zr, $^{90}$Mo, $^{88}$Nb, $^{89}$Nb (dashed) and $^{89}$Mo (dash-dotted), respectively (see text)}
\label{natMod_d,xnxp_ALL}
\end{figure*}

\section{Conclusions}
\label{Sum}

A consistent analysis has concerned all reaction mechanisms involved in the complex deuteron interaction process, e.g. elastic scattering, breakup, stripping, pick-up, pre-equilibrium and compound nucleus, including the contributions of specific Mo isotopes too. 
The brief overview of the activation data for the deuteron--induced reactions on $^{nat}$Mo, the related TENDL--2021 evaluation \cite{T21}, and model calculations carried out within present work is shown in Figs.~\ref{natMod_d,xn} and \ref{natMod_d,xnxp_ALL}. 
It is thus underlined comparatively the neutron emission $^{nat}$Mo$(d,xn)$ and the neutron emission possibly accompanied by H and He isotopes emission $^{nat}$Mo$(d,xnyp)$, if emerging nucleon clusters are given due consideration.  
They are related to the high requests of the radioactivity risks and radiation damages estimation, critical benchmark for selecting and validating the best structural materials and a number of key technologies. 
Thus, accurate estimation of $(d,xn)$ activation is essential for shielding design of nuclear installations while H, T, and He gas emission leading to "gas bubbles accumulation", are of high interest for damage studies of the structural materials. 
Overall, the due account of most experimental data has thoroughly validated the present theoretical framework and highlighted some prevalent features. 

Hence strong neutron emission goes through $(d,2n)$ and $(d,3n)$ reaction channels (Figs.~\ref{99mTc_natMod}--\ref{93Tc_natMod}), while the charged--particle emission (protons, tritons, $^{3,4}$He) is almost an order of magnitude less (Fig.~\ref{natMod_d,xnxp_ALL}). 
There is only one different case in this respect, of the strong protons emission within $^{100}$Mo$(d,p)^{101}$Mo reaction. 
Then, higher protons and even tritons emission leads to $^{99}$Mo residual nucleus, with a maximum of $\sim$65 mb for the corresponding excitation function below 10 MeV [Fig.~\ref{natMod_d,xnxp_ALL}(b)]. 
Increased emission of neutrons, protons, $^3$He and $\alpha$-particles concern the activation of Nb and Zr isotopes, whose estimation is important for both theory validation and structural materials comparative analysis. 

At the same time, this comparison has emphasized the weak points and consequently the need for modeling/evaluation upgrade. 
Most of them were related to overlooking the deuteron inelastic--breakup enhancement and appropriate treatment of stripping and pick-up processes shown in Sec.~\ref{Activation}. 
Nevertheless, there may still be present particular discrepancies that are related to the complexity of the interaction process, not entirely accounted for in routine evaluation/theoretical analyzes. 
New data as well as complementary measurements of $(d,px)$ and $(n,x)$ as well as $(d,nx)$ and $(p,x)$ reaction cross sections for isotopes of the natural element, presently very scarce, are essential for validation of the theoretical frame associated to the deuteron breakup process. 
The final improvement of evaluation predictions becomes thus feasible for target nuclei and incident energies where data are missing, so important for advanced engineering design projects. 


\section*{Acknowledgments}

This work has been partly supported by OP RDE, MEYS, Czech Republic under the project SPIRAL2-CZ, CZ.02.1.01/0.0/0.016\_013/0001679 and  by The Executive Unit for the Financing of Higher Education, Research, Development and Innovation (UEFISCDI) under the project No. PN-III-ID-PCE-2021-0642), and carried out within the framework of the EUROfusion Consortium, funded by the European Union via the Euratom Research and Training Programme (Grant Agreement No 101052200 — EUROfusion). Views and opinions expressed are however those of the author(s) only and do not necessarily reflect those of the European Union or the European Commission. Neither the European Union nor the European Commission can be held responsible for them.











\appendix

\section{Spectroscopic factors}\label{secA1}
\subsection{Table content}
The spectroscopic factors obtained through the DWBA analysis of the experimental outgoing--particle angular distributions (Figs. 3--8) are given hereafter. 
The data in the following tables concern the excitation energies $E$ of the residual--nucleus states (assumed by authors of the angular--distribution measurements), their spin $J$ and parity $pi$, the transferred angular momentum $L$, and the corresponding spectroscopic factor $S$ obtained in the present work by analysis of the experimental outgoing--particle angular distributions shown in Figs. 3--8. 
The spectroscopic factors that have been considered for calculation of the rest of stripping and pickup reactions were taken from the given references in the text.
On the other hand, the same deuteron OMP \cite{dah} as well as the nucleon \cite{KD}, triton \cite{BG}, and $\alpha$-particle \cite{avr2014} OMPs have been involved within DWBA calculations as within BU+PE+CN models, for a consistent model analysis.




\subsubsection{$^{92}$Mo$(d,n)^{93}$Tc [Fig. 3(top)]}
\begin{verbatim}
----------------------------
E(MeV)  J	  pi 	 L	   S
----------------------------
0.00	  4.5	  1  	4	  0.40
0.39	  0.5	 -1  	1	  0.14
1.50	  1.5	 -1  	1	  0.045
1.78	  0.5	 -1  	1	  0.075
2.59	  2.5	  1  	2	  0.035
3.17	  2.5	  1   2	  0.035
3.36	  2.5	  1	  2	  0.115
3.90	  0.5	  1	  0	  0.032
4.11	  0.5	  1	  0	  0.3
4.76	  2.5	  1	  2	  0.21
4.90	  1.5	  1	  2	  0.125
5.01	  2.5	  1	  2	  0.15
5.18	  2.5	  1	  2	  0.23
5.30	  2.5	  1	  2	  0.26
5.44	  0.5	  1	  0   0.3
5.62	  0.5	  1	  0   0.195
5.68	  0.5	  1	  0   0.15
5.78	  2.5	  1	  2   0.22
5.93	  0.5	  1	  0   0.27
-----------------------------
\end{verbatim} 

\subsubsection{$^{94}$Mo$(d,n)^{95}$Tc [Fig. 3(bottom)]}
\begin{verbatim}
----------------------------
E(MeV)	  J	   pi	 L	   S
----------------------------
0.000	  4.5	  1	  4	  0.38
0.040	  0.5	 -1	  1	  0.235
0.064	  1.5	 -1	  1	  0.105
1.084	  2.5	  1	  2	  0.026
1.276	  1.5	  1	  2	  0.08
1.281	  3.5	 -1	  3	  0.06
1.450	  2.5	  1	  2	  0.065
1.639	  1.5	 -1	  1	  0.025
1.750	  1.5	 -1	  1	  0.035
2.320	  2.5	  1	  2	  0.035
2.550	  0.5	 -1	  1	  0.06
2.830	  2.5	  1	  2	  0.145
3.210	  0.5	  1	  0	  0.045
3.490	  2.5	  1	  2	  0.035
3.630	  2.5	  1	  2	  0.04
3.810	  0.5	  1	  0	  0.035
3.920	  0.5	  1	  0	  0.02
3.990	  1.5	  1	  2	  0.031
4.400	  1.5	  1	  2	  0.023
4.500	  1.5	  1	  2	  0.033
1.660	  0.5	  1	  0	  0.04
-----------------------------
\end{verbatim} 
\subsubsection{$^{95}$Mo$(^{3}He,d)^{96}$Tc [Fig. 4(top)]}
\begin{verbatim}
-------------------------------
E(MeV)	  J	   pi	  L	    S
-------------------------------
0.000	   5	   1	   4	   0.600		 		
0.046	   5	   1	   4	   0.780		 
0.121	   2	  -1	   1	   0.275		 
0.177	   4	  -1	   1	   0.150		 
0.227	   3	  -1	   1	   0.130		 
0.228	   4	   1	   4	   0.292		  
0.316	   3	  -1	   1	   0.210		 
0.319	   6	   1	   4	   0.300		 		
0.352	   3	  -1	   1	   0.215		 
0.506	   5	   1	   4	   0.235		 
0.568	   3	  -1	   1	   0.090		 
0.623	   4	   1	   4	   0.201		 
0.750	   3	  -1	   1	   0.085		 
0.801	   4	  -1	   1	   0.100		 
0.815	   4	   1	   4	   0.201		 
0.867	   4	  -1	   1	   0.090		 
0.933	   3	   1	   2	   0.120		 
0.980	   5	   1	   2	   0.077		 
1.066	   2	   1	   2	   0.130		 
1.158	   4	  -1	   1	   0.083		 
1.255	   4	  -1	   1	   0.095		 
1.314	   5	   1	   2	   0.115		 
1.338	   5	   1	   2	   0.110		 
1.408	   4	  -1	   1	   0.125		 
1.482	   4	  -1	   1	   0.095		 
1.536	   5	   1	   2	   0.090		 
1.597	   4	  -1	   1	   0.120		 
1.653	   4	  -1	   1	   0.040		 
1.670	   5	   1	   2	   0.080		 
1.772	   5	   1	   2	   0.115		 
1.825	   5	   1	   2	   0.085		 
1.884	   5	   1	   2	   0.080		 
1.940	   5	   1	   2	   0.074		 
---------------------------------
\end{verbatim} 
\subsubsection{$^{96}$Mo$(^{3}He,d)^{97}$Tc (Fig. 4)}
\begin{verbatim}
------------------------------
E(MeV)	 J	    pi	  L	   S
------------------------------
0.000	  4.5	   1	  4	  0.600		 
0.096	  0.5	  -1	  1	  0.350		 
0.326	  2.5	   1	  2	  0.080		 
0.576	  1.5	  -1	  1	  0.180		 
0.655	  2.5	  -1	  3	  0.110		 
0.783	  2.5	   1	  2	  0.170		 
0.947	  1.5	  -1	  1	  0.145		 
1.053	  1.5	  -1	  1	  0.065		 
1.316	  4.5	   1	  4	  0.180		 
1.374	  2.5	   1	  2	  0.065		 
1.537	  0.5	   1	  0	  0.065		 
1.599	  1.5	   1	  2	  0.120		 
1.649	  2.5	   1	  2	  0.100		 
1.712	  0.5	   1	  0	  0.110		 
1.847	  0.5	   1	  0	  0.160		 
1.951	  1.5	   1	  2	  0.200		 
2.013	  2.5	   1	  2	  0.085		 
2.111	  1.5	   1	  2	  0.085		 
2.151	  0.5	   1	  0	  0.075		 
2.264	  0.5	   1	  0	  0.140		 
2.307	  2.5	   1	  2	  0.090		 
2.653	  0.5	   1	  0	  0.170		 
2.713	  1.5	   1	  2	  0.120		 
2.783	  2.5	   1	  2	  0.070		 
2.878	  1.5    1	  2	  0.085		 
2.908	  2.5	   1	  2	  0.100		 
3.018	  1.5	   1	  2	  0.100		 
3.060	  2.5	   1	  2	  0.110		 
3.145	  0.5	   1	  0	  0.160		 
3.214	  0.5	   1	  0	  0.105		 
------------------------------
\end{verbatim} 
\subsubsection{$^{98}$Mo$(^{3}He,d)^{99}$Tc (Fig. 5)}
\begin{verbatim}
------------------------------
E(MeV)	 J	    pi	  L	   S
------------------------------
0.000	  4.5	   1	  4	  0.460		 
0.142	  0.5	  -1	  1	  0.215		 
0.181	  2.5	   1	  2	  0.040		 
0.509	  1.5	  -1	  1	  0.130		 
0.625	  3.5	   1	  4   0.160		 
0.672	  2.5	  -1	  3	  0.090		 
0.720	  3.5	   1	  4	  0.107		 
0.762   2.5	   1	  2	  0.110		 
0.919	  0.5	   1	  0	  0.045		 
1.020	  1.5	   1	  2	  0.055		 
1.203	  0.5	  -1	  1	  0.050		 
1.321	  0.5	  -1	  1	  0.070		 
1.435	  1.5	   1	  2	  0.080		 
1.505	  1.5	   1	  2	  0.053		 
1.560	  0.5	   1	  0	  0.130		 
1.679	  1.5	   1	  2	  0.080		 
1.760	  1.5	   1	  2	  0.062		 
1.803	  0.5	   1	  0	  0.115		 
1.825	  1.5	   1	  2	  0.100		 
1.911	  1.5	   1	  2	  0.065		 
1.982	  1.5	   1	  2	  0.110		 
2.000	  1.5	   1	  2	  0.105		 
2.111	  2.5	   1	  2	  0.100		 
2.160	  1.5	   1	  2	  0.100		 
2.176	  0.5	   1	  0	  0.050		 
2.203	  1.5	   1	  2	  0.085		 
2.281	  0.5	   1	  0	  0.110		 
2.396	  1.5	   1	  2	  0.065		 
2.414	  1.5	   1	  2	  0.060		 
2.466	  0.5	   1	  0	  0.065		 
2.486	  0.5	   1	  0	  0.065		 
2.522	  1.5	   1	  2	  0.085		 
2.581	  1.5	   1	  2	  0.095		 
2.653	  0.5	   1	  0	  0.060		 
2.675	  0.5	   1	  0	  0.060		 
------------------------------
\end{verbatim} 
\subsubsection{$^{92}$Mo$(d,\alpha)^{90}$Nb (Fig. 6)}
\begin{verbatim}
----------------------------------
E(MeV)	 J	  pi	  L	     S
----------------------------------
0.000 	  8	  1	  8	    0.6	 
0.129 	  4	 -1	  3	    0.75 
0.176*	  7   1	  6+8	  0.8+0.1
0.286*	  5	  1	  4+6   0.4+0.05	 
0.381 	  1	  1	  0	    0.65 
0.655 	  3	  1	  2	    0.365	 
0.819 	  9	  1	  8	    2.2 
0.823 	  2	 -1	  1	    0.48 
0.848 	  2	  1	  2	    0.65	 
0.958 	  4	 -1	  3	    0.26	 
1.131 	  5	 -1	  5	    0.36	 
1.194 	  2	 -1	  1	    0.11	 
1.231*	  4	 -1	  3+5	  0.25+0.1
1.255*	  4	 -1	  3+5	  0.15+0.1	 
1.289*	  4	 -1	  3+5	  0.08+0.12	  
1.350 	  1	  1	  0	    0.25	 
1.370*	  2	 -1	  1+3	  0.36+0.24	 
1.415 	  4	 -1	  3	    0.8	 
1.492*	  4	 -1	  3+5	  1.2+0.3	 
1.552*	  4	 -1	  3+5	  0.7+0.75 
1.647 	  4	 -1	  3	    0.4	 
1.691 	  4	 -1	  3	    0.35	 
1.776*	  1	  1	  0+2	  0.35+0.1	 
1.842*	  1	  1	  0+2	  0.18+0.2	
1.873 	  3	 -1	  3	    0.32	 
2.000 	  4	 -1	  3	    0.38 
2.138 	  1	  1	  0	    0.31 
2.311*	  1	  1	  0+2	  0.18+0.05 
----------------------------------
*Two possible L values considered.
\end{verbatim} 
\subsubsection{$^{92}$Mo$(d,t)^{91}$Mo [Fig. 7(top)]}
\begin{verbatim}
-----------------------------
E(MeV)  J	   pi 	 L 	  S
-----------------------------
0.000	  4.5	  1	  4	  2.96
0.652	  0.5	 -1	  1	  1.29
1.155	  1.5	 -1	  1	  1.4
1.364	  2.5	  1	  2	  0.436
1.533	  2.5	 -1	  3	  1.61
1.904	  4.5	  1	  4	  0.93
2.085	  1.5	 -1	  1	  0.387
2.241	  2.5	  1	  2	  0.15
2.299	  1.5	 -1	  1	  0.292
2.462	  2.5	 -1	  3	  0.387
2.547	  2.5	  1	  2	  0.188
2.727	  2.5	 -1	  3	  0.806
2.824	  4.5	  1	  4	  0.755
2.898	  1.5	 -1	  1	  0.31
3.188	  1.5	 -1	  1	  0.21
3.330	  1.5	 -1	  1	  0.243
3.455	  1.5	 -1	  1	  0.346
2.302	  4.5	  1	  4	  0.316
3.351	  2.5	 -1	  3	  0.387
-----------------------------
\end{verbatim} 

\subsubsection{$^{94}$Mo$(d,t)^{93}$Mo  (Fig. 7)} 
\begin{verbatim}
-----------------------------
E(MeV)   J	  pi 	 L 	  S
-----------------------------
0.000	  2.5	  1	  2	  1.35
0.947	  0.5	  1	  0	  0.43
1.364	  3.5	  1	  4	  0.6
2.529	  0.5	 -1	  1	  0.69
2.534	  4.5	  1	  4	  1.7
3.510	  4.5	  1	  4	  0.78
3.587	  4.5	  1	  4	  1.2
3.590	  1.5	 -1	  1	  0.387
3.650	  4.5	  1	  4	  0.83
3.720	  1.5	 -1	  1	  0.4
4.630	  1.5	 -1	  1	  0.4
4.710	  1.5	 -1	  1	  0.265
4.720	  2.5	 -1	  3	  0.65
4.756	  1.5	 -1	  1	  0.2
4.780	  4.5	  1	  4	  0.43
5.000	  1.5	 -1	  1	  0.4
5.034	  1.5	 -1	  1	  0.43
5.070	  4.5	  1	  4	  0.5
5.150	  1.5	 -1	  1	  0.32
------------------------------
\end{verbatim} 
\subsubsection{$^{98}$Mo$(d,t)^{97}$Mo  (Fig. 7)}
\begin{verbatim}
-----------------------------
E(MeV)  J	  pi 	 L 	  S
-----------------------------
0.00	  2.5	  1	  2	  1.15
0.68	  0.5	  1	  0	  0.42
0.72	  1.5	  1	  2	  0.26
0.89	  0.5	  1   0	  0.182
1.12	  2.5	  1	  2	  0.24
1.28	  2.5	  1	  2	  0.48
2.39	  0.5	 -1	  1	  0.47
2.52	  4.5	  1	  4	  0.82
2.83	  0.5	 -1	  1	  0.3
------------------------------
\end{verbatim} 
\subsubsection{$^{100}$Mo$(d,t)^{99}$Mo (Fig. 7)}
\begin{verbatim}
-----------------------------
E(MeV)    J	  pi 	 L 	  S
-----------------------------
0.000	   0.5	  1	  0	  0.45
0.099	   2.5	  1	  2	  0.44
0.236	   3.5	  1	  4	  0.87
0.351	   1.5	  1	  2	  0.07
0.524	   0.5	  1	  0	  0.155
0.548	   1.5	  1	  2	  0.65
0.617	   2.5	  1	  2	  0.55
0.684	   5.5	 -1	  5	  0.76
0.697	   3.5	  1	  4	  0.33
0.756	   3.5	 -1	  3	  0.43
0.797	   1.5	  1	  2	  0.29
0.894	   1.5	  1	  2	  0.23
0.912	   0.5	  1	  0	  0.31
0.951	   2.5	  1	  2	  0.43
1.030	   1.5	 -1	  1	  0.17
1.051	   2.5	 -1	  3	  0.23
1.148	   3.5	 -1	  3	  0.14
1.173	   2.5	  1	  2	  0.22
1.201	   1.5	  1	  2	  0.21
1.258	   2.5	  1	  2	  0.12 
1.352	   4.5	  1	  4	  0.31
1.497	   2.5	  1	  2	  0.255
1.545	   2.5	  1	  2	  0.3
1.580	   1.5	  1	  2	  0.21
1.893	   2.5	 -1	  3	  0.305
1.944	   0.5	  1	  0	  0.195
2.160	   4.5	  1	  4	  0.925
2.220	   2.5	 -1	  3	  0.81
2.482	   0.5	  1	  0	  0.15
2.517	   4.5	  1	  4	  0.9
------------------------------
\end{verbatim} 
\subsubsection{$^{96}$Mo$(d,^{3}He)^{95}$Nb (Fig. 8)}
\begin{verbatim}
------------------------------
E(MeV)	   J	  pi	  L	   S
------------------------------
0.000	   4.5	  1	  4	  2.050		 
0.232	   0.5	 -1	  1	  1.150		 
0.792	   1.5	 -1	  1	  0.950		 
1.000	   2.5	 -1	  3	  1.320		 
1.221	   1.5	 -1	  1	  1.200		 
1.623	   1.5	 -1	  1	  0.380		 
1.635	   2.5	  1	  2	  0.250		 
1.980	   2.5	 -1	  3	  0.650		 
2.230	   1.5	 -1	  1	  0.250		 
2.328	   2.5	 -1	  3	  0.750		 
2.340	   1.5	 -1	  1	  0.450		 
2.481	   2.5	 -1	  3	  1.050		 
2.786	   2.5	 -1	  3	  1.050		 
------------------------------
\end{verbatim} 

\subsubsection{$^{98}$Mo$(d,^{3}He)^{97}$Nb  (Fig. 8)}
\begin{verbatim}
----------------------------
E(MeV)   J	  pi 	 L 	  S
----------------------------
0.000	   4.5	  1	  4	  2.150		 
0.746	   0.5	 -1	  1	  1.300		 
1.251	   1.5	 -1	  1	  1.800		 
1.438	   2.5	 -1	  3	  2.300		 
1.764	   1.5	 -1	  1	  0.400		 
1.774	   2.5	  1	  2	  0.250		 
2.090	   2.5	  1	  2	  0.450		 
2.100	   2.5	 -1	  3	  1.200		 
2.244	   1.5	 -1	  1	  0.600		 
2.386	   1.5	 -1	  1	  0.650		 
2.550	   1.5	 -1	  1	  0.650		 
2.948	   1.5	 -1	  1	  0.600		 
2.963	   2.5	 -1	  3	  0.900		 
----------------------------
\end{verbatim} 




\end{document}